# Structural and lattice-dynamical properties of Tb₂O₃ under compression: a comparative study with rare-earth and related sesquioxides

Jordi Ibáñez,[1] Juan Ángel Sans,[2] Vanesa Cuenca-Gotor,[2] Robert Oliva,[3*] Óscar Gomis,[4] Plácida Rodríguez-Hernández,[5] Alfonso Muñoz,[5] Ulises Rodríguez-Mendoza,[5] Matías Velázquez,[6] Philippe Veber,[7] Catalin Popescu[8] and Francisco Javier Manjón[2*]

[1]*Institute of Earth Sciences Jaume Almera, MALTA Consolider Team, Consell Superior d'Investigacions Científiques (CSIC),
08028 Barcelona, Catalonia, Spain*
[2]*Instituto de Diseño para la Fabricación y Producción Automatizada, MALTA Consolider Team, Universitat Politècnica de València, 46022 València, Spain*
[3]*Faculty of Fundamental Problems of Technology, Wroclaw University
of Science and Technology, 50-370, Wrocław, Poland*
[4]*Centro de Tecnologías Físicas, MALTA Consolider Team, Universitat Politècnica de València, 46022 València, Spain*
[5]*Departamento de Física, Instituto de Materiales y Nanotecnología, MALTA Consolider Team, Universidad de La Laguna, 38200 San Cristóbal de la Laguna, Tenerife, Spain*
[6]*Univ. Grenoble Alpes, CNRS, Grenoble INP, SIMAP, 38000 Grenoble, France*
[7]*CNRS, Institut Lumière Matière, Université Claude Bernard Lyon 1, UMR5306, 69622 Villeurbanne, France*
[8]*ALBA-CELLS, MALTA Consolider Team, 08290 Cerdanyola del Vallès (Barcelona), Catalonia, Spain.*

*Corresponding authors: robert.oliva.vidal@pwr.edu.pl; fjmanjon@fis.upv.es

**Abstract**

We report a joint experimental and theoretical investigation of the high-pressure structural and vibrational properties of terbium sesquioxide (Tb₂O₃). Powder x-ray diffraction and Raman scattering measurements show that cubic $Ia\bar{3}$ (C-type) Tb₂O₃ undergoes two phase transitions up to 25 GPa. We observe a first irreversible reconstructive transition to the monoclinic $C2/m$ (B-type) phase at ~7 GPa and a subsequent reversible displacive transition from the monoclinic to the trigonal $P\bar{3}m1$ (A-type) phase at ~12 GPa. Thus, Tb₂O₃ is found to follow the well-known C→B→A phase transition found in other cubic rare-earth sesquioxides with cations of larger atomic mass than Tb. Our *ab initio* theoretical calculations predict phase transition pressures and bulk moduli for the three phases in rather good agreement with experimental results. Moreover, Raman-active modes of the three phases have been monitored as a function of pressure and lattice-dynamics calculations have allowed us to confirm the assignment of the experimental phonon modes in the C-type phase as well as to make a tentative assignment of the symmetry of most vibrational modes in the B- and A-type phases. Finally, we extract the bulk moduli and the Raman-active modes frequencies together with their pressure coefficients for the three phases of $Tb_2O_3$. These results are thoroughly compared and discussed in relation to those reported for rare-earth and other related sesquioxides. It is concluded that the evolution of the volume and bulk modulus of all the three phases of these technologically relevant compounds exhibit a nearly linear trend with respect to the third power of the ionic radii of the cations.

**1.- Introduction**

Rare-earth (RE) sesquioxides (SOs), in particular lanthanide SOs ($Ln_2O_3$; $Ln$=La to Lu, Y and Sc), are an important family of materials due to their remarkable fundamental properties and potential applications. These compounds are highly interesting and versatile for different types of applications because the Ln radius can be finely tuned along the lanthanide family with the filling of *f* orbitals, thus enabling a wide range of technological advances including light emitters (lasers and improved phosphors), catalysts, and high-dielectric constant (high-k) gates. In particular, terbium sesquioxide ($Tb_2O_3$) has attracted considerable attention in the last few years as a high-k material[1–3] and also as an active material for optical insulators and high-performance optoelectronic devices.[4–6]

It is well known that RE SOs exhibit three polymorphic modifications at room conditions depending on the RE radius: i) a trigonal phase, named A-type (space group (s.g.) $P\bar{3}m1$, No. 164, Z=1), for large cations from La to Nd; ii) a monoclinic phase, named B-type (s.g. $C2/m$, No. 12, Z=6), for medium-size cations from Sm to Gd; and iii) a cubic bixbyite-type phase, named C-type (s.g. $Ia\bar{3}$, No. 206, Z=16), for small cations from Tb to Lu, including Y and Sc. The bixbyite phase also occurs at ambient pressure in other SOs, like $Mn_2O_3$, $In_2O_3$ and $Tl_2O_3$. At high temperatures, two additional phases named as H (hexagonal, s.g. $P6_3/mmc$, No. 194, Z=2) and X (cubic, s.g. $Im\bar{3}m$, No. 229, Z=4) have also been found in RE SOs.[7] Phase transitions (PTs) of $Ln_2O_3$ at high temperatures have been summarized by Zinkevich[8] and many of them usually follow the sequence C→B→A on increasing temperature. In particular, $Tb_2O_3$ was observed to undergo the C→B PT above 1600ºC-1875 ºC,[7,9,10] followed by the B→H PT above 2430-2450ºC[7,11,12] and the H→X PT above 2610-2640ºC.[11,12]

Since the molar volume (density) of the A-, B- and C-type phases decreases (increases) in the order C, B, A, the PT sequence C→B→A is expected for most RE SOs crystallizing in the cubic phase, either at high pressure (HP) or at high temperature (HT).[13,14] The effect of HP on RE SOs has been extensively studied by several research groups and their pressure-induced PTs have been recently reviewed.[15] In fact, since the cubic phase can be synthesized in a number of them, from Lu to Sm as well as in $Mn_2O_3$, $In_2O_3$ and $Tl_2O_3$, its behavior under compression has been widely studied, as for instance in $Lu_2O_3$,[16,17] $Yb_2O_3$,[18,19] $Tm_2O_3$,[20,21] $Er_2O_3$,[22–24] $Ho_2O_3$,[25–28] $Dy_2O_3$,[29,30] $Gd_2O_3$,[31–38] $Eu_2O_3$[39–45] $Sm_2O_3$,[34,46,47] $Sc_2O_3$,[48–50] $Y_2O_3$,[34,51–61] $Mn_2O_3$,[62–65] $In_2O_3$,[66–70] and $Tl_2O_3$.[71] Moreover, the equation of state of several RE SOs of C-, B- and A-type has been studied comparatively.[72]

Polymorphism of RE SOs and their transformations at HP have also been investigated theoretically and the simulations of the bandgap, volume, bulk modulus and its pressure derivative for the C-, B- and A-type phases as well as their PT pressures have been reported and found in rather good agreement with experiment.[73–85] In fact, several theoretical works have reported the study of the bulk modulus of C-type RE SOs as a function of the atomic number;[84,86] however, the overall behavior of the bulk modulus as a function of unit cell volume and cation radius remains to be investigated for all three C-, B- and A-type phases. Besides, the explanation of the behavior of the bulk modulus as a function of the unit-cell volume for all C-type SOs, including not only $Ln_2O_3$ but also bixbyite-type SOs, such as $Sc_2O_3$, $Mn_2O_3$, $Y_2O_3$, $In_2O_3$ and $Tl_2O_3$, is still awaiting.

Despite the large amount of work performed on the study of $Ln_2O_3$ properties at HP, relatively little is known about $Tb_2O_3$ under compression. Hoekstra and co-workers transformed the cubic phase into the monoclinic one at 2.5 GPa and 905 ºC.[13] However, no experimental data are available, up to our knowledge, for A-type $Tb_2O_3$ to compare with theoretical estimations.[76,79,84,86] The lack of studies on $Tb_2O_3$ lies on the difficulty of synthesizing this material that usually crystallizes at room conditions in the C-type structure. The reason for the difficult synthesis of $Tb_2O_3$ is the tendency of terbium oxide to grow naturally as $Tb_4O_7$, a mixed valence compound with Tb atoms in both 3+ and 4+ valence states.[75,87] Fortunately, it has been recently shown that pure millimeter-sized C-type $Tb_2O_3$ single crystals can be grown with a controlled atmosphere flux method.[88] The high-quality cubic $Tb_2O_3$ crystals thus obtained are highly promising for applications in optics and photonics. Using such high-quality material, the room-pressure lattice-dynamics properties of cubic $Tb_2O_3$ have been recently investigated.[89]

In the last two decades, a vast number of high-pressure studies has been devoted to investigate the HP behavior of RE SOs. Although the pressure-induced phase transition sequence has been well stablished, the structural and vibrational properties of most of these compounds under compression; i.e., the equation of state (bulk moduli) and vibrational properties are not so well understood and show little consistency across the investigated literature. Therefore, a more consistent work from both experimental and theoretical point of views is required to achieve a better understanding of the structural and vibrational properties of RE SOs.

In this work, we present a thorough experimental and theoretical study on the structural and vibrational properties of $Tb_2O_3$ under compression, where powder x-ray diffraction (XRD) and Raman scattering (RS) measurements at HP have been complemented with theoretical *ab initio* calculations. We will show that C-type $Tb_2O_3$ undergoes two PTs up to 25 GPa: a first irreversible C→B PT, and a second reversible B→A PT. The experimental and theoretical equation of state (EoS) of the three phases of $Tb_2O_3$ will be provided and compared with isostructural compounds. In particular, we will show that there is a linear trend between the bulk moduli of the three C-, B-, and A-type phases and the cube cation ionic radius. Additionally, up to thirteen optical Raman-active modes of C-type $Tb_2O_3$ have been monitored as a function of pressure, while sixteen and four Raman-active modes have been observed in the B- and A-type phases, respectively. The assignment of the symmetries of the different phonons in the three phases is supported by the results of *ab initio* lattice-dynamics calculations, and the experimental and theoretical phonon pressure coefficients in the three phases have been compared with those of isostructural SOs.

## 2.- Experimental methods

Millimeter-sized pure cubic $Tb_2O_3$ single crystals with [112] crystallographic orientation were the starting material employed in this study.[88] Details of the growth method and the structural and vibrational properties of the $Tb_2O_3$ single crystals at room conditions can be found in Refs. [88,89]. The single crystals were grinded for powder XRD experiments, which were first performed at room conditions with a PANalytical X'Pert Pro MPD diffractometer (Cu Kα1 Kα2 radiation). Le Bail fits confirmed the cubic $Ia\bar{3}$ symmetry and allowed extracting a lattice parameter, $a$ = 10.7323(3) Å, and volume, $V_0$=1236.17 Å$^3$, in good agreement with previous structural characterizations.[10,90] No reflections arising from the $Li_6Tb(BO_3)_3$ solvent employed during growth, or from other impurity phases, were detected in the XRD scans.

Powder angle-dispersive XRD measurements at HP were performed at room temperature in the BL04-MSPD beamline at ALBA synchrotron facility.[91] This beamline is equipped with Kirkpatrick-Baez mirrors to focus the monochromatic beam and a Rayonix CCD detector with a 165 mm diameter active area. The XRD measurements were performed with monochromatic x-rays with a wavelength of 0.4246 Å, as determined from the absorption K-edge of Sn (29.2 keV). The sample was loaded with a 4:1 methanol-ethanol mixture in a membrane-type diamond anvil cell (DAC) with diamond culets of 400 μm in diameter. The applied pressure was determined with the EoS of copper.[92] The associate error to the pressure determination, including the pressure gradients inside the DAC, is lower than 0.5 GPa. The sample to detector distance (240 mm), along with various detector geometrical parameters of the experiment, were calibrated with the DIOPTAS software[93] using diffraction data from $LaB_6$ measured in the same conditions as the sample. The structural analyses were performed with the program TOPAS 4.2. For the low-pressure (LP) cubic phase, Rietveld refinements to the XRD patterns were carried out, whereas for the HP phases the lattice parameters were obtained with Pawley/Le Bail whole pattern fittings. **Figure S1** in the Supplementary Information (SI) shows selected examples of calculated and difference profiles as obtained from the profile refinements.

Room-temperature unpolarized HP-RS measurements were excited with the 532-nm line of a solid state laser. For these experiments, micron-size fragments without well-known orientation were obtained from the single crystals. Inelastically-scattered light was collected with a Horiba Jobin Yvon LabRAM HR UV spectrometer equipped with an edge filter that cuts Raman signals below ~50 cm$^{-1}$ and a thermoelectrically cooled multichannel CCD detector enabling a spectral resolution better than 2 cm$^{-1}$. The applied pressure was determined using the luminescence of small ruby chips evenly distributed in the pressure chamber.[94] Phonon signals were analyzed by fitting the Raman peaks with a pseudo-Voigt profile.

## 3.- First-principles calculations

*Ab initio* total-energy calculations at 0 K for the C-, B-, and A-type phases of Tb$_2$O$_3$ (and also for other SOs) were performed within the framework of density functional theory (DFT)[95] with the Vienna Ab-initio Simulation Package (VASP),[96] using the pseudopotential method and the projector augmented waves (PAW) scheme.[97,98] In this work, the generalized gradient approximation (GGA) with the Perdew-Burke-Ernzerhof (PBE) parametrization[99] was used for the exchange and correlation energy, after evaluating calculations performed also with its extension to the solid state (PBEsol).[100] A dense Monkhorst−Pack grid[101] of special *k*-points (6×6×6 for the C phase and 4×4×4 for the A and B phases) and a plane-wave basis set with energy cutoffs of 530 eV were used. Lattice-dynamical properties were obtained for the Γ-point using the direct-force constant approach,[102] with both PBE and PBEsol functionals in order to compute the atomic forces. A comparison of data with both GGA-PBE and GGA-PBEsol prescriptions led us to use, in most cases, the GGA-PBE data for the joint analysis with the experimental results, unless otherwise specified.

## 4.- Results and discussion

### 4.1.- Structural properties of Tb$_2$O$_3$ under pressure

Tb$_2$O$_3$ crystallizes in the cubic C-type polymorph (or bixbyite-type), a structure traditionally understood as derived from the fluorite lattice by doubling the lattice parameter and leaving one-fourth of the anion sites vacant in an ordered way (see **Fig. 1**). In fact, the bixbyite structure is an intermediate structure between the zincblende and the fluorite structure, since the cation array, which derives from the face-centered cubic structure present in *Ln* atoms at ambient conditions, is the same in the three structures.[103,104]

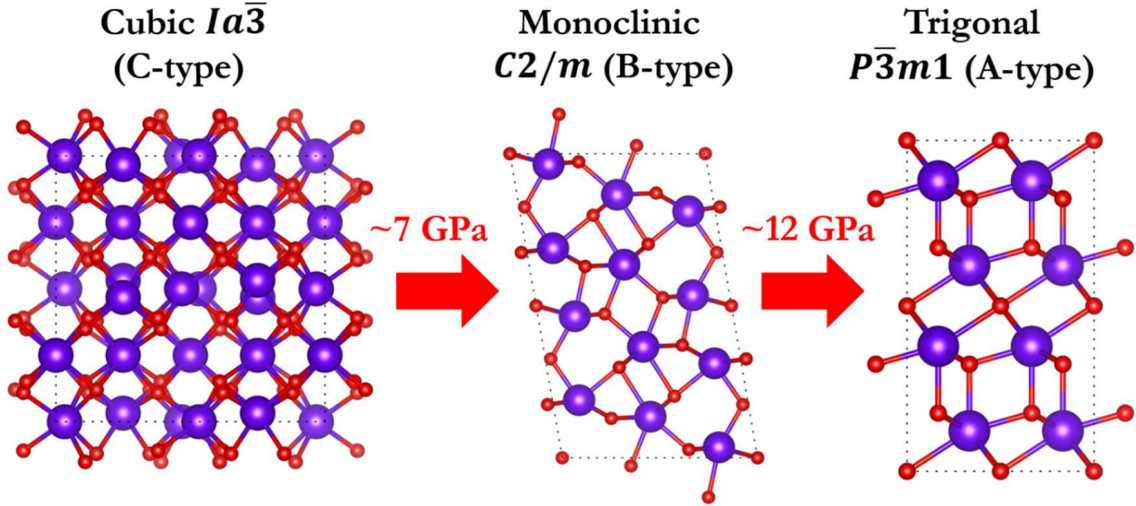

**Figure 1:** Crystal structure of the C, B and A phases of $Tb_2O_3$. Structural phase transitions take place at ~7 GPa (C→B) and ~12 GPa (B→A).

In C-type $Tb_2O_3$ there are three independent atoms, Tb1, Tb2 and O. Tb1 and Tb2 atoms are located at *8b* and *24d* Wyckoff sites, respectively, while O atoms occupy *48e* sites. Therefore, this structure has only 4 free atomic coordinates, the $x_{Tb}$ parameter of the Tb2 atom, usually denoted as *u*, and the three atomic coordinates ($x_O, y_O, z_O$) of the O atom. In this structure, O atoms show 4-fold coordination and both Tb atoms show 6-fold coordination. The differences in bonding found for the two different *Ln* atoms of the C-type structure have been previously discussed.[105] In particular, O atoms around Tb1 form a regular octahedron with 6 equal Tb-O distances around 2.285 Å, while O atoms around Tb2 form an irregular octahedron with three different Tb-O distances around 2.274, 2.297 and 2.391 Å. The different environment of the two inequivalent Tb sites can be understood if one considers that the C-type structure derives from the fluorite lattice, where two O atoms out of the eight O atoms surrounding Tb atoms in a cube are removed along (111) and (110) directions.[80] This gives rise to two different configurations for Tb atoms, being the first configuration more regular than the second one.

**Figure 2** shows the HP-XRD patterns of $Tb_2O_3$ at room temperature for selected pressures up to 20 GPa. It can be observed that all the scans from 0 to ~7 GPa are dominated by peaks of the C-type phase. However, new reflections, which start to be visible at 7.1 GPa (not seen in **Fig. 2** due to scaling reasons) and are observed up to 11.2 GPa, can be indexed with the B-type phase.

On the upstroke, the B-type phase was always found to be accompanied by the C-type or the A-type phase, this last phase emerging around 12.0 GPa. Above this pressure, the A-type peaks start to broaden upon compression, most probably as a consequence of the loss of hydrostatic conditions of the pressure-transmitting medium.[106] The experimental values of 7.1 and 12.0 GPa for the C→B and B→A PTs, respectively, are in good agreement with those predicted by our calculations using GGA-PBE DFT calculations, which locate both PTs around 5.5 and 10.9 GPa, respectively (see **Figure 3**). These values are also in good agreement with the phase transition pressures reported in other RE SOs.[15] The XRD peaks of the A-type phase show up in the scans up to the maximum pressure of the experiment (20 GPa). Although these reflections could also be indexed with the H-type phase, RS measurements (see below) confirm that the phase above 12.0 GPa corresponds to the A-type structure. On the downstroke, this phase was found to back-transit to the B-type phase around 10 GPa, being this latter phase metastably observed down to room pressure (see the top XRD scan in **Fig. 2**). In fact, the volume and lattice parameters obtained for the B-type phase at room pressure (see **Tables 1 and 2**) agree nicely with those previously reported. [10,107,108]

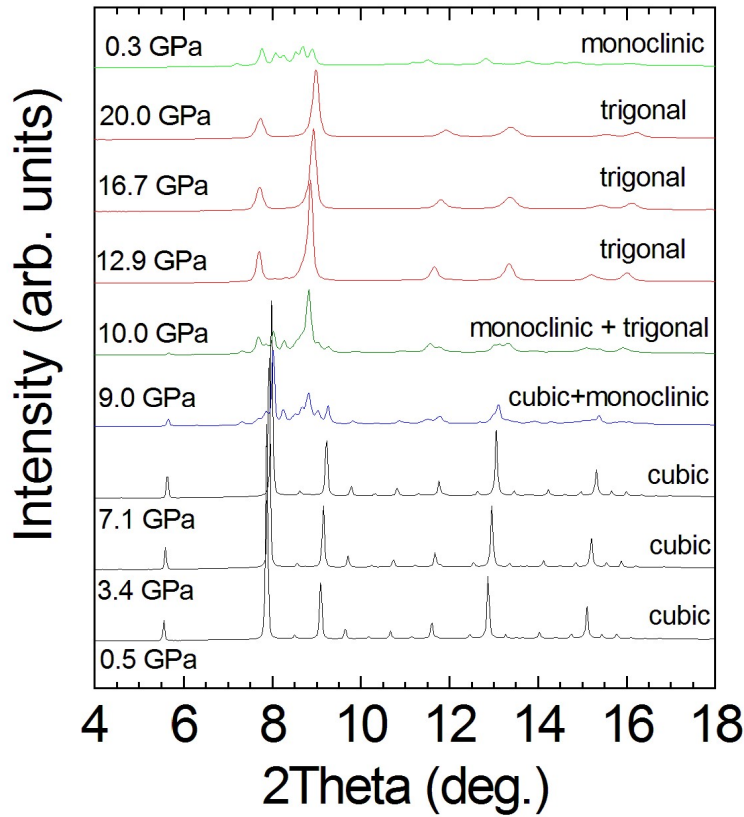

**Figure 2:** Room-temperature HP-XRD patterns of $Tb_2O_3$ at selected pressures up to 20 GPa. The top scan shows the resulting pattern at 0.3 GPa after pressure release in the downstroke.

**Table 1**: EoS parameters of the cubic (C), monoclinic (B), and trigonal (A) phases of $Tb_2O_3$ obtained from fits to the experimental (exp) and DFT-PBE theoretical (the) data obtained in this work. The last column indicates the type of EoS employed to fit the data. Volumes are given per formula unit.

| Phase | $V_0$ (Å$^3$) | $B_0$ (GPa) | $B_0'$ | EOS |
|---|---|---|---|---|
| C (exp) | 77.3(1) | 148(2) | 2.1(4) | BM3 |
| C (the) | 77.5(1) | 140.5(4) | 4.2(1) | BM3 |
| B (exp) | 71.0(1) | 133(3) | 4 (fixed) | BM2 |
| B (the) | 71.5(2) | 135.3(6) | 4 (fixed) | BM2 |
| A (exp) | 71.0(6) | 112(5) | 4 (fixed) | BM2 |
| A (the) | 70.3(1) | 133.9(2) | 4 (fixed) | BM2 |

**Table 2**: Experimental (exp) and GGA-PBE theoretical (the) zero-pressure lattice parameters of the cubic (C) and monoclinic (B) phases of $Tb_2O_3$. The parameters of the trigonal (A) phase of $Tb_2O_3$ are given at 12.2 GPa.

|  | $a$ (Å) |  | $b$ (Å) |  | $c$ (Å) |  | $\beta$ (deg.) |  |
|---|---|---|---|---|---|---|---|---|
|  | Exp. | The. | Exp. | The. | Exp. | The. | Exp. | The. |
| C-type [a] | 10.732(1) | 10.743 |  |  |  |  |  |  |
|  | 10.730[c] | 10.735[d] |  |  |  |  |  |  |
| B-type [a] | 14.005(5) | 14.136 | 3.543(5) | 3.535 | 8.718(5) | 8.718 | 100.1(1) | 100.33 |
|  | 14.04(1)[b] | 14.13[e] | 3.541(3)[b] | 3.537[e] | 8.725(8)[b] | 8.716[e] | 100.06(5)[b] | 100.21[e] |
| A-type [a] | 3.658(5) | 3.646 |  |  | 5.610(5) | 5.646 |  |  |
|  |  | 3.715[e] |  |  |  | 5.858[e] |  |  |

[a] This work; [b] Ref. [107]; [c] Ref. [109]; [d] Ref. [86]; [e] Ref. [79].

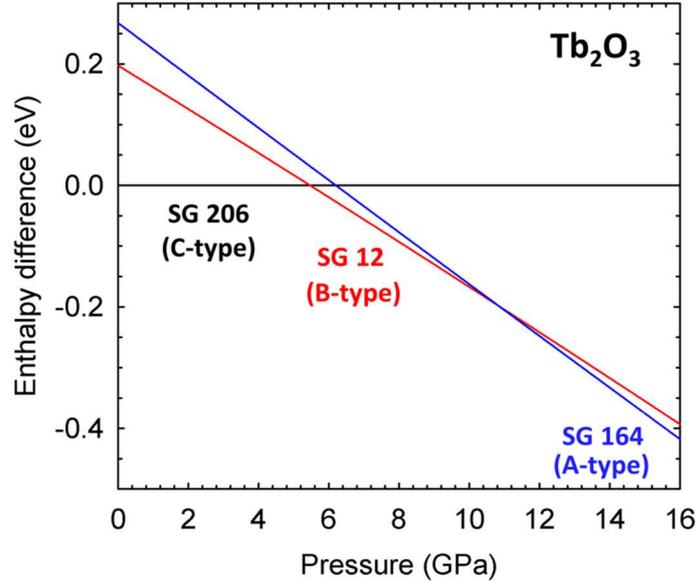

**Figure 3:** Enthalpy difference between the B- (red line) and A-type (blue line) phases of $Tb_2O_3$ with respect to the C-type (black line) phase as a function of pressure.

The experimental pressure dependence of the unit-cell volume of the different phases of $Tb_2O_3$ up to 20 GPa, as obtained from Rietveld analysis, is plotted in **Figure 4**. For comparison purposes, all the data in this figure correspond to the unit-cell volume per formula unit (pfu). Theoretical data are also plotted in **Fig. 4** with solid lines for direct comparison with the experimental results. A fit of pressure vs. experimental unit-cell volume to a 3$^{rd}$-order Birch-Murnaghan (BM) EoS[110] yields, for the case of C-type $Tb_2O_3$ ($Z$=16), a zero-pressure unit-cell volume, $V_0$, bulk modulus, $B_0$, and pressure derivative of the bulk modulus, $B_0$', of $V_0$=1236.2(2) Å$^3$, $B_0$=148(2) GPa and $B_0$'=2.1(4), respectively. The value of $V_0$ obtained from the 3$^{rd}$-order BM-EoS fit is similar to that measured outside the DAC and in previous reports.[10,90] We have also fitted the experimental data with a 2$^{nd}$-order BM-EoS, which yields $V_0$=1237.0(2) Å$^3$ and a somewhat larger bulk modulus $B_0$=139(2) GPa ($B_0$'=4 in this case). On the other hand, the fit of theoretical GGA-PBE data using a 3$^{rd}$-order BM-EoS yields the following values: $V_0$=1240.12(4) Å$^3$, $B_0$=140.5(4) GPa and $B_0$'=4.2(1), respectively. It must be noted that there is an excellent agreement between our experimental and theoretical $V_0$ and $B_0$ values for C-type $Tb_2O_3$ (see **Table 1**). Although the experimental bulk modulus obtained with the 2$^{nd}$-order EoS is somewhat closer to the theoretical value, here we rely on the result obtained with the 3$^{rd}$-order EoS because the normalized stress vs Eulerian strain plot (not shown) is slightly better in this case. Note, on the other hand, that our theoretical bulk modulus for C-type $Tb_2O_3$ is also in good agreement with some recent theoretical predictions using GGA-PBE calculations[84,86] (see **Tables S1 and S2** in the SI, which display experimental and theoretical data for different sesquioxides). The theoretical and experimental values of $V_0$, $B_0$ and $B_0$' for the C-type phase, as well as those for the B-type and A-type phases of $Tb_2O_3$ studied below, are summarized in **Table 1**; whereas experimental and theoretical data of the lattice parameters for the three phases can be found in **Table 2**.

The good agreement observed between our theoretical and experimental data for the unit-cell volume of C-type $Tb_2O_3$ is also found in the evolution of its free atomic coordinates (see **Fig. S2a**) and, therefore, in the different Tb-O distances as a function of pressure. Results of theoretical calculations and of Rietveld refinement of the experimental data for $Tb_2O_3$ show that the four free atomic coordinates $x_{Tb}$, $x_O$, $y_O$, and $z_O$ exhibit a negligible change between 0 and 10 GPa (**Fig. S2a**). This result implies that there is a monotonous decrease of the Tb-O distance with pressure in the C-type phase (**Fig. S2b** in the SI) that is basically determined by the lattice parameter reduction.

We have not found in the literature the report of any pressure dependence of the atomic coordinates or of the cation-anion interatomic distances in other isostructural C-type or bixbyite-type SOs. Our present calculations in $Lu_2O_3$, $Sc_2O_3$, and $In_2O_3$ (see **Fig. S3**) show that their atomic parameters also exhibit a negligible change with pressure irrespective of the high-pressure phase acquired by the cubic SO. The negligible change of atomic parameters and the consequent monotonous and parallel decrease of interatomic distances of cubic SOs under hydrostatic compression is very interesting. On one hand, it means that there is no net increase of the effective coordination number either of the metal atoms or of O atoms on increasing pressure and, on the other hand, that the two low-symmetry Wyckoff positions for metal (24*d*) and O (48*e*) atoms in the C-type/bixbyite-type structure behave like very high-symmetry positions; i.e., with all atomic coordinates fixed. This behavior is in contrast to what happens in corundum-type SOs, like $In_2O_3$ and $Al_2O_3$ (see **Fig. S4**). As observed, the theoretical free atomic parameters of $Al_2O_3$ up to 30 GPa show the opposite behavior than those of $In_2O_3$. Clearly, the different change of the free atomic parameters of the corundum phase in $Al_2O_3$ and $In_2O_3$ explains why $InO_6$ and $AlO_6$ octahedra evolve in a different way with pressure, as recently reported.[70]

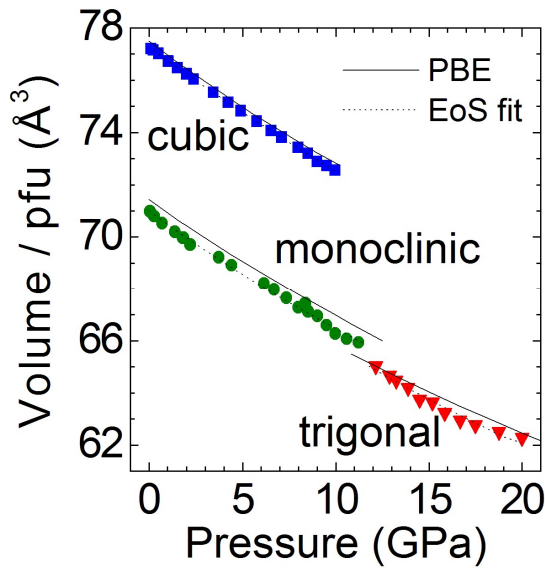

**Figure 4:** Volume per formula unit as a function of the applied pressure for the three phases of $Tb_2O_3$ as obtained experimentally (symbols). Solid lines represent the results of *ab initio* calculations. Calculated errors for the experimental values are smaller than the sizes of the symbols in all cases.

We have to note that the negligible change of the atomic parameters of the C-type structure of SOs with pressure is consistent with the negligible variation of those parameters upon changing ionic radius, as noted by Richard *et al.*[83] In this context, it must be recalled that the lanthanide contraction on going from La to Lu and even to Sc can be viewed as a chemical pressure similar to physical pressure, so both the effects of pressure and contraction of ionic radius are comparable. **Figure S3** shows that the atomic coordinates of non-lanthanide bixbyite-type SOs ($In_2O_3$, $Tl_2O_3$, $Mn_2O_3$, etc.) are exactly the same as those of C-type RE SOs (including $Sc_2O_3$ and $Y_2O_3$), so both types of compounds are isostructural. This adds additional interest to the different pressure-induced phase transitions observed in these isostructural compounds, since the cubic lanthanide SOs undergo transitions to B- and A-type phases while cubic non-lanthanide SOs undergo transitions to phases related to the corundum-type structure.

The C→B PT occurs in Tb$_2$O$_3$ at ~7 GPa on the upstroke. This is a first-order PT with an experimental (theoretical) relative volume reduction of 8% (7.8%). The B-type phase, shown in **Fig. 1**, is characterized by having one O atom at a *2b* site as well as three Tb atoms and four O atoms at *4i* Wyckoff sites. This structure has no fixed atomic coordinates, so the seven inequivalent atoms at *4i* sites contains 21 free coordinates. In particular, B-type Tb$_2$O$_3$ has two Tb atoms (Tb1 and Tb2) with sevenfold coordination and one Tb atom (Tb3) with sixfold coordination. As regards O atoms, O1 has fivefold coordination, O2, O3 and O4 have fourfold coordination, and O5 has sixfold coordination. Therefore, on average the coordination of Tb is around 6.5 and that for O is around 4.8. The B-type phase of RE SOs is also characterized by a cationic array that is a distortion of the hexagonal closed packed (hcp) structure of metals (P6$_3$/mmc, No. 194, Z=2). In the B-phase, O atoms are located in voids with tetrahedral arrangement within the hcp structure.[111] Notably, Tb-Tb distances are 3.278, 3.536, 3.657 and 4.114 Å in B-type Tb$_2$O$_3$ at room pressure,[108] giving an average Tb-Tb distance of 3.646 Å which is only slightly larger (3.4%) than the Tb-Tb distance in hcp Tb (3.525 Å).[112]

Full pattern matching of B-type Tb$_2$O$_3$ allowed us to extract its lattice parameters at different pressures between 7 and 11 GPa on the upstroke and also from 10 GPa down to room pressure on the downstroke. As expected, a monotonous decrease of the unit-cell volume of B-Tb$_2$O$_3$ up to 11 GPa is observed (**Fig. 4**). Fit of pressure vs. experimental unit-cell volume pfu (Z=6) to a 2$^{nd}$-order BM-EoS yields $V_0$=426.1(5) Å$^3$ and $B_0$=133(3) GPa, respectively. In this case, we rely on the 2$^{nd}$-order EoS because third-order terms are not found to better explain the normalized stress vs Eulerian strain dependence for this phase (not shown). The results thus obtained are in very good agreement with results of our theoretical GGA-PBE calculations fit to a 2$^{nd}$-order BM-EoS that yields $V_0$= 428.8(1) Å$^3$ and $B_0$= 135.3(6). It can be observed that the room-pressure unit-cell volume, lattice parameters and monoclinic β angle obtained in this work for B-type Tb$_2$O$_3$ agree nicely with those obtained by other authors (see **Tables 1 and 2**).[10,107,108] Additionally, the bulk modulus is similar to that found in other monoclinic RE SOs, like Eu$_2$O$_3$,[113] but smaller than that of Y$_2$O$_3$.[114]

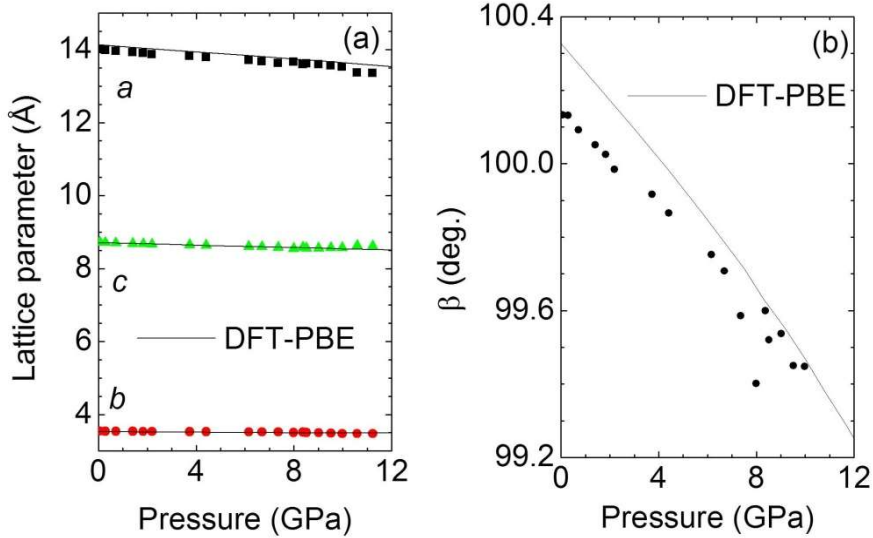

**Figure 5:** Experimental (symbols) and theoretical (lines) pressure dependence of the lattice parameters and β monoclinic angle of B-type $Tb_2O_3$. The calculated errors for the refined lattice parameters and monoclinic β angle are smaller, in all cases, than symbol sizes.

The experimental lattice parameters of B-type $Tb_2O_3$ up to 12 GPa (see **Fig. 5a**) display a monotonous and smooth decrease as pressure increases in good agreement with our theoretical GGA-PBE calculations. Similarly, the experimental and theoretical pressure dependence of its axial ratios $c/b$, $a/b$ and $a/c$ (see **Fig. S5** in the SI) are also found to display a monotonous behavior that is well reproduced by our calculations. On the other hand, a small reduction of the monoclinic β angle (see **Fig. 5b**), always perpendicular to the $b$-axis, with increasing pressure is found. This result implies that in B-type $Tb_2O_3$ the direction of the $a$-axis displays a slight change with pressure (assuming that the directions of both $b$ and $c$ axis remain constant). All these observations and the good agreement between the GGA-PBE calculations and experimental results support the present theoretical description of the compression of B-type $Tb_2O_3$. The analysis of the theoretical and experimental compressibility tensors of this phase (see Tables S9 and S10 of the **SI material**) shows that the $a$-axis of the monoclinic cell is more compressible than the $b$ and $c$ axes regardless of pressure. On the other hand, the theoretical pressure dependence of the Tb-O distances in B-type $Tb_2O_3$ is plotted in **Fig. S6** in the SI. With the exception of one of the $Tb_1$-O and $Tb_3$-O distances, there is a monotonous decrease of the Tb-O distances with pressure similar to that found for the C-type phase. This plot reflects the complex rearrangement of the different atoms of the monoclinic cell upon compression, which ends up with the B→A PT above 10 GPa (see **Fig. 2**). Indeed, the B→A PT is of weak first-order character since it shows a very small volume reduction (<2% according to GGA-PBE).

Trigonal A-type $Tb_2O_3$, shown in Fig. 1, is characterized by having one Tb atom at a *2d* site, one O atom (O1) at a *2d* site and one O atom (O2) at a *1a* Wyckoff site. Therefore, this structure has only two free atomic coordinates corresponding to the *z* atomic positions of the two atoms at *2d* sites. In A-type $Tb_2O_3$, Tb is sevenfold coordinated, O1 is fourfold coordinated and O2 has sixfold coordination. This means that on average the coordination of O atoms is 5, very similar to that of the B-type structure, while the coordination of Tb is slightly larger than in the B-type structure. This result is in agreement with the A-type phase $P\bar{3}m1$ being related to the B-type structure (which is a distortion of the former). Therefore, it can be concluded that the A-type phase is also characterized by a cation array similar to the hcp structure of metals, as the B-type phase.[111] Note that the B-type structure derives from the A-type phase by a slight lattice deformation implying a splitting of *1a* ($D_{3d}$) and *2d* ($C_{3v}$) atomic positions into less symmetrical *2b* ($C_{2h}$) and *4i* ($C_s$) sites, as commented by Gouteron *et al.*[115]

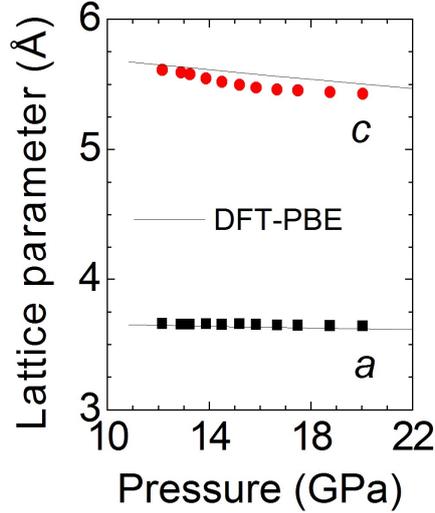

**Figure 6:** Experimental (symbols) and theoretical (lines) pressure dependence of the lattice parameters of A-type $Tb_2O_3$. The calculated errors for the experimental lattice parameters are smaller than the symbol sizes.

Whole pattern matching of A-type $Tb_2O_3$ XRD scans allowed us to get the lattice parameters of this phase on upstroke from 12 to 20 GPa. As expected, a monotonous decrease of the unit-cell volume pfu (Z=1) of A-type $Tb_2O_3$ with pressure is observed (**Fig. 4**). Fit of pressure vs. experimental unit-cell volume to a 2$^{nd}$-order BM-EoS yields $V_0$= 71.0(6) Å$^3$ and $B_0$= 112(5) GPa. As already commented, there is no previous experimental published data in the literature for this phase of $Tb_2O_3$. It must be stressed that, again, our theoretical GGA-PBE data closely follow the experimental results, fit to a 2$^{nd}$-order BM-EoS yields $V_0$=70.3(1) Å$^3$ and $B_0$= 134(1) GPa. As observed, our theoretical bulk modulus is slightly larger than the experimental one. The difference between them will be later commented. In this context, we can mention that the experimental bulk modulus of $Tb_2O_3$ is similar to that found in $Ce_2O_3$[116] and smaller than that of $Nd_2O_3$.[117]

The pressure dependence of the lattice parameters of A-type $Tb_2O_3$ is reported in **Fig. 6**. As observed, there is a monotonous decrease of both *a* and *c* lattice parameters of the trigonal phase with increasing pressure and there is a good agreement between the experimental and theoretical results. It must be noted that the experimental (theoretical) value of the *c/a* ratio for A-type $Tb_2O_3$ at 12.2 GPa is 1.534 (1.548) according to data in **Table 2**. In turn, the theoretical value expected at 0 GPa for A-type $Tb_2O_3$ is 1.577, while that of A-type $Y_2O_3$ is of 1.617.[79] It has been suggested that the *c/a* ratio of A-type RE SOs from $La_2O_3$ to $Sm_2O_3$ tends to the ideal value of 1.633 that corresponds to the hexagonal close packing of cations.[115] However, such a value is not observed in any theoretical estimation of the RE SOs, including A-type $Y_2O_3$ and A-type $Sc_2O_3$. In fact, all theoretical *c/a* values of A-type RE SOs (including $Y_2O_3$) remain below 1.62, except for $Sc_2O_3$ that is around 1.67.[79] Therefore, it seems that the relationship between the distortion of the hexagonal close packing of cations and the ionic radius seems not to be a valid idea for A-type RE SOs.

**Figure S7** in the SI shows the pressure dependence of the *c/a* axial ratio in A-type $Tb_2O_3$. As can be seen, both experiment and theory indicate that the *c/a* ratio decreases with pressure; i.e. upon compression it progressively deviates from the ideal value of the hexagonal close packing of cations. The experimental curve, however, exhibits a kink around 15 GPa that is probably a consequence of the broad XRD reflections that are detected for this phase, giving rise to large experimental errors in the determination of the lattice parameters. **Figures S8a and 8b** in the SI show the pressure dependence of the theoretical free atomic coordinates and Tb-O distances in A-type $Tb_2O_3$. As observed, the GGA-PBE calculations predict almost no change in the two free atomic parameters ($z_{Tb}$, $z_O$) with increasing pressure as well as a monotonous decrease of Tb-O distances, with no sudden changes that could give rise to the kink observed in the experimental *c/a* ratio. As it was already commented for the case of the C-type phase, it can be concluded that the decrease of Tb-O distances with pressure in A-type $Tb_2O_3$ is fully determined by the lattice parameter reduction due to negligible change in the free atomic parameters with pressure. We want to note that the change of the free atomic parameters with increasing pressure in B-type $Tb_2O_3$ (not shown) is also rather small although not negligible. These results are consistent with the negligible change of theoretical free atomic parameters of both B-type and A-type phases on changing ionic radius.[79] In fact, the experimental free atomic parameters of B-type $Sc_2O_3$ at 38.5 GPa[48] are similar to the theoretical ones reported for $Sc_2O_3$ at 0 GPa by Wu *et al*.[79] and at 15.7 GPa by Zhang *et al*.[118] In this context, we must remind that a value of 0.5 must be added to experimental values of the free coordinate *x* of Sc atoms and of O(4) and O(5) atoms in Ref. [48]

(due to a lattice origin shift) for a proper comparison with the theoretical values plotted in **Fig. S8a**.

**4.2.- Compression behavior of RE and related sesquixodes**

Next, we would like to compare the bulk moduli of A-, B- and C-type $Tb_2O_3$ with those of other RE SOs and isostructural SOs. Only a few works have attempted to understand the overall compression behavior of this family of materials.[84,86] For this purpose, we have summarized in **Table S1** in the SI the experimentally reported values of the bulk moduli of some of these compounds in the C-type/bixbyite structure. As discussed in Ref. [84], experimental data show considerable dispersion, making almost impossible to draw any conclusion about the behavior of the bulk moduli of the C-type SOs. To address this issue, here we have performed theoretical DFT-PBE calculations for several C-type SOs, including $Sc_2O_3$, $Y_2O_3$, $In_2O_3$, $Sm_2O_3$, $Tb_2O_3$, $Lu_2O_3$, and $Tl_2O_3$ corresponding to atomic numbers 21, 39, 49, 61, 65, 71 and 81. Moreover, we have also performed the same calculations for $Al_2O_3$, $Fe_2O_3$, $Ga_2O_3$, $V_2O_3$, $Mn_2O_3$, $Cm_2O_3$, $Pu_2O_3$, and $La_2O_3$ in order to extend the range of cation ionic radii in possible C-type/bixbyite SOs. **Figure S9a** in the SI shows a rather good agreement of our theoretical and reported experimental unit-cell volume pfu in almost all C-type SOs, with an upward sawtooth-like tendency with increasing the atomic number. Contrarily, our theoretical bulk moduli show a systematic slightly underestimation of the corresponding experimental data (**Fig. S9b** in the SI), but both the experimental and theoretical values show a clear overall tendency to decrease with increasing Z. This result suggests that the reduction of the bulk modulus (i.e., increase of the compressibility) with Z is in fact related to the corresponding unit-cell volume increase. Opposite to the overall tendency, the region corresponding to the RE SOs shows a decrease (increase) of the volume (bulk modulus) with increasing Z, in good agreement with previous calculations.[80,86]

It can be observed that our theoretical GGA calculations slightly overestimate the experimental volumes, being the largest differences found in $In_2O_3$ (~5%) and $Tl_2O_3$ (~7%). The lanthanide contraction effect is well reproduced by our calculations and agrees well with previous works.[73,75,80,84,86] Noteworthy, the largest deviations in the theoretical vs. experimental bulk modulus also correspond to $In_2O_3$ (~22%) and $Tl_2O_3$ (~28%). The large deviations for these two compounds, observed on the light of the comparison between their theoretical unit-cell volumes and their well-known experimental unit-cell volumes in **Fig. S9a**, suggest that the theoretical values are clearly underestimated, but also that the experimental bulk moduli are likely quite overestimated. Similarly, the comparison of experimental reported values of the bulk modulus and our theoretical estimations for RE SOs suggest that the experimental bulk modulus reported for $Dy_2O_3$, $Gd_2O_3$ and $Yb_2O_3$ seems to be quite overestimated whereas that of $Lu_2O_3$ seems to be underestimated.

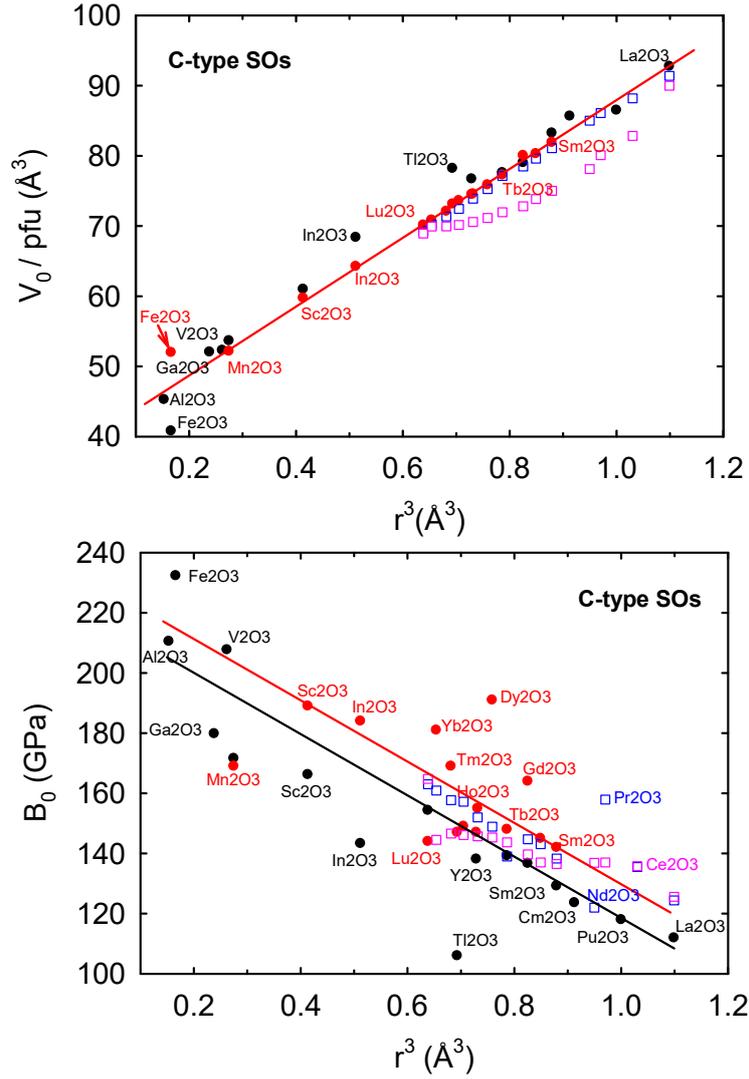

**Figure 7:** Experimental (red) and theoretical (black) unit-cell volume per formula unit **(a)** and bulk modulus **(b)** of C-type sesquioxides as a function of the cube cation ionic radius. Theoretical data of Ref. [84] with WC-GGA (pink open squares) and GGA+U (blue open squares) are also shown for comparison. Red lines show the linear trend in experimental values, while black line suggest the linear trend in theoretical bulk modulus according to our calculations. Cation ionic radii have been taken from Ref. [119] for valence 3+ with 6-fold coordination.

In **Figure 7a** we have plotted the experimental and theoretical unit-cell volume pfu as a function of the third power of the cation ionic radius, $r^3$, in C-type SOs. It can be observed that there is a clear linear trend between both parameters (see red circles and red line), as already stablished by Shannon and Prewitt in 1970.[120] This linear dependence is a direct consequence of the negligible change of the free atomic parameters of the C-type structure on changing the cation ionic radius, as already discussed. Moreover, the linear dependence between unit-cell volume and $r^3$ implies that there cannot be a linear relationship between the unit-cell volume of C-type SOs and the cation ionic radius, as initially suggested by Hoekstra in 1966.[107] In fact, our theoretical calculations for the C-type phase show a relationship between the unit-cell volume and cation ionic radii (not shown) similar to that reported for B- and A-type phases in Ref. [79]. In this context, we must stress that our calculations are more comparable to GGA+U calculations for RE SOs than to WC-GGA ones.[84] Moreover, our calculations allow us to extend the observed trend even farther than previously reported (from $Al_2O_3$ to $La_2O_3$). As observed, the largest deviation of theoretical values from experimental values are those of cubic $In_2O_3$, $Tl_2O_3$ and $Fe_2O_3$. This suggests that GGA-PBE is not the best method to describe the unit-cell volume for these three compounds. Furthermore, our graph suggests that the experimental unit-cell volume reported for C-type $Fe_2O_3$[121] either does not follow the linear trend of other C-type SOs or it is considerably overestimated. In fact, our calculations estimate the unit-cell volume per formula unit of C-type $Fe_2O_3$ to be around 47.32 Å$^3$.

In **Fig. 7b** we show the experimental and theoretical bulk modulus as a function of $r^3$ in C-type SOs. Again, a linear trend can be plotted between both parameters (see red circles and red line). Note that our calculated values display rather good agreement with previous ones.[84] The negative slope of the bulk modulus with the third power of the cation ionic radius clearly shows the inverse relationship between the bulk modulus and the unit-cell volume. Thus, with the exception of the lanthanides due to the contraction effect, the increase of cation Z leads in general to larger cation ionic radius, larger unit-cell volumes and smaller bulk moduli. Our graph shows that our calculated bulk moduli are somewhat underestimated as compared to experimental ones. Similar to the case of the unit-cell volume, the largest underestimation corresponds to $In_2O_3$ and $Tl_2O_3$. Similarly, the bulk modulus of C-type $Fe_2O_3$ is expected to be somewhat overestimated and a value close to 220 GPa is expected. Unfortunately, there is no reported bulk modulus, to our knowledge, for this metastable phase of $Fe_2O_3$.

Finally, it is worth noting that similar conclusions to those here obtained for the C-type/bixbyite phase of SOs can be reached for the B-and A-type phases of SOs. **Figures S10a and S11a** in the SI show the increase of the unit-cell volume pfu in B- and A-type RE SOs with increasing Z, except along the lanthanide series due to the lanthanide contraction effect. Our calculations are in good agreement with other theoretical calculations.[80,84,86] Contrarily, the bulk modulus shows the opposite trend and decreases with increasing Z, except along the lanthanide series. As observed, our experimental and theoretical values for the bulk moduli of B- and A-type $Tb_2O_3$ are within the error margin, and it seems that the experimental value of the bulk modulus of the A-type phase is slightly underestimated.

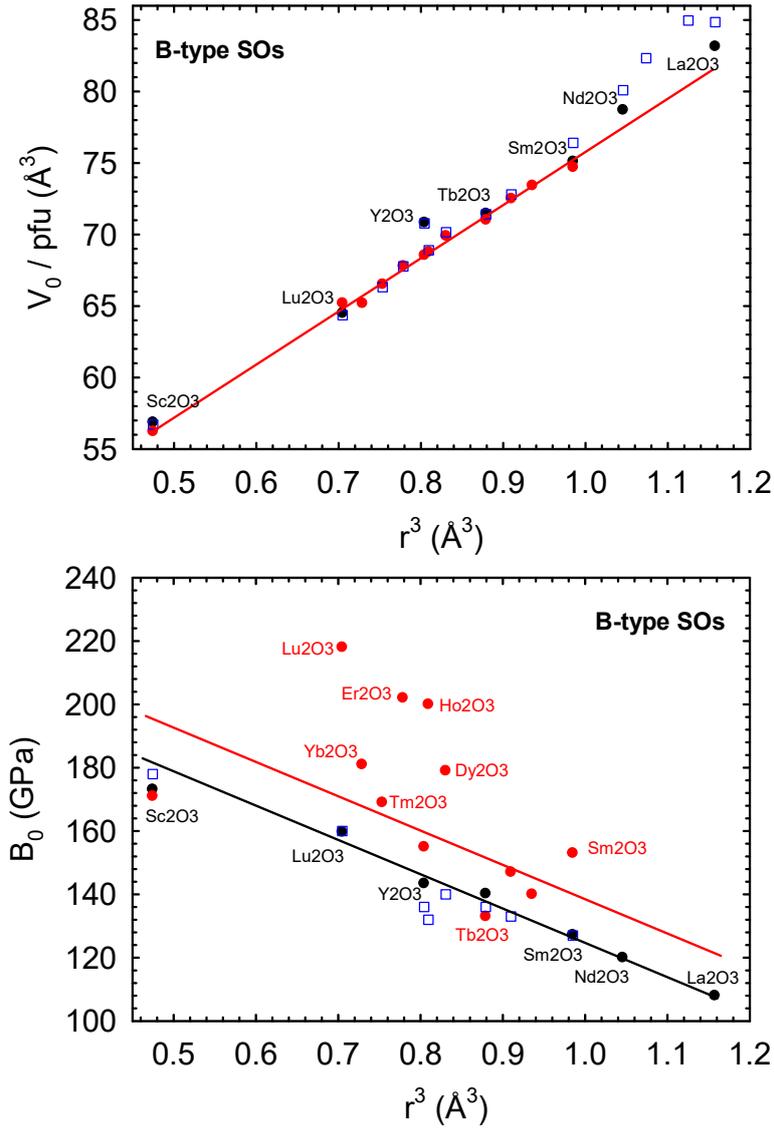

**Figure 8:** Experimental (red) and theoretical (black) unit-cell volume per formula unit **(a)** and bulk modulus **(b)** of B-type sesquioxides as a function of the third power of the cation ionic radius ($r^3$). Theoretical data of Ref. [79] (blue open squares) are also shown for comparison. Red lines show the linear trend in experimental values, while the black line indicates the linear trend in theoretical bulk modulus according to our calculations. Cation ionic radii have been taken from Ref. [119] for valence 3+ with 6.5-fold coordination.

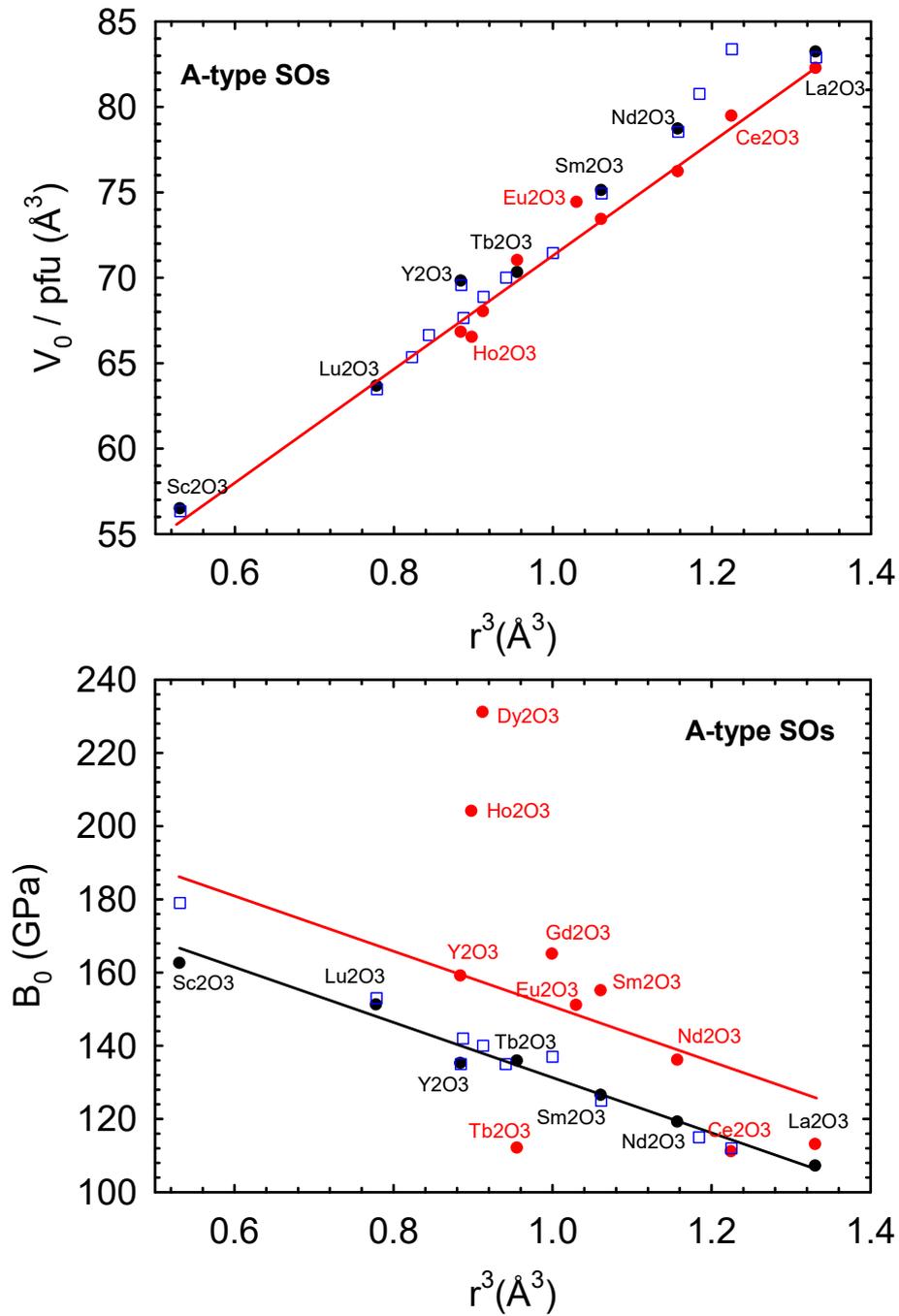

**Figure 9:** Experimental (red) and theoretical (black) unit-cell volume per formula unit **(a)** and bulk modulus **(b)** of A-type sesquioxides as a function of the third power of the cation ionic radius ($r^3$). Theoretical data of Ref. [79] (blue open squares) are also shown for comparison. Red lines show the linear trend in experimental values, while the black line indicates the linear trend in theoretical bulk modulus according to our calculations. Cation ionic radii have been taken from Ref. [119] for valence 3+ with 7-fold coordination.

**Figures 8a and 9a** show the theoretical (GGA-PBE) unit-cell volume pfu as a function of $r^3$ for some RE SOs, including $Sc_2O_3$ and $Y_2O_3$, with B- and A-type structure, respectively. Data for other SOs (with filled d orbitals) are not included in the plot because these materials do not exhibit these two phases. It can be observed that there is a linear relationship between both parameters in both phases. Again, the linear behavior between unit-cell volume of the B- and A-type structures and the third power of the cation ionic radius is a direct consequence of the negligible change of the free atomic parameters in both structures with the cation ionic size, as already discussed. We have to stress that both experiments and calculations show that there is no linear relationship between the unit-cell volume of B-type and A-type SOs and the cation ionic radius, as initially suggested by Hoekstra.[107] This conclusion is especially clear when data from $Sc_2O_3$ to $La_2O_3$ are included to extend the range of ionic radii. In this context, our theoretical calculations compare rather well with calculations for B- and A-type RE SOs.[79] As observed, the largest deviation of theoretical values from experimental values are those for $Y_2O_3$.

In turn, **Figures 8b and 9b** show the theoretical (GGA-PBE) bulk moduli as a function of $r^3$ for some RE SOs, including $Sc_2O_3$ and $Y_2O_3$, with B- and A-type structures, respectively. As can be seen in both figures, there is a linear relationship between the bulk modulus and $r^3$ for both B- and A-type structures that suggest an almost inverse relationship between the bulk modulus and the unit-cell volume for the A- and B-type phases of the RE SOs, including $Sc_2O_3$ and $Y_2O_3$. Moreover, the comparison of theoretical and experimental data may suggest that the values of the bulk moduli of B-type $Lu_2O_3$, $Er_2O_3$, $Dy_2O_3$, and $Ho_2O_3$ seem to be quite overestimated and the same holds for A-type $Dy_2O_3$ and $Ho_2O_3$, while those of A-type $Tb_2O_3$ and $Ce_2O_3$ are likely underestimated. We have to note that our results on the linear behavior of the bulk modulus of B- and A-type SOs with respect to the cation ionic radius is in agreement with previous calculations.[79] Finally, it must be noted that our theoretical values for the bulk modulus of the B-type (A-type) SOs are in general slightly higher (lower) to those for C-type SOs. This is also in agreement with theoretical results of Ref. [79] but in disagreement with those of Ref. [84], which reports a bulk modulus of A-type SOs 10-15% higher than that of C-type SOs. On the other hand, our theoretical bulk moduli of SOs are larger for the B-type phase than for the A-type phase. This result is also in agreement with previous theoretical estimations.[76]

In summary, we have proved that our theoretical calculations for different C-, B- and A-type SOs reproduce the well-known lanthanide unit-cell volume contraction with the increase of the atomic number as well as the inverse relationship between the bulk modulus and the unit-cell volume (or the third power of the cation ionic size) for the three phases. Consequently, we can conclude that the compressibility of the C-, B- and A-type phases of RE SOs is mainly determined by the unit-cell volume and correspondingly by the cation ionic radii. In this way, we have shown a general overview of the behavior of RE SOs under compression that complements the one recently reported.[15] Moreover, we have extended the discussion regarding the linear dependence of the unit-cell volume and bulk moduli of RE SOs (including $Sc_2O_3$ and $Y_2O_3$) vs the cation ionic size to other C-type/bixbyite-type SOs. On top of that, we have found that our experimental bulk moduli for the different phases of $Tb_2O_3$ are within the error margin of measurements, likely being the bulk modulus of A-type $Tb_2O_3$ slightly underestimated. Besides, we have pointed out that several experimental bulk moduli of SOs must be revised since they seem to be quite overestimated.

**4.3.- Vibrational properties of $Tb_2O_3$ under pressure: comparison to related sesquioxides**

RS spectroscopy is a powerful nondestructive analytical tool that provides useful information, among others, about the crystal quality, structural properties and lattice dynamics of solid-state materials. Many different studies have been devoted to investigate the vibrational properties of C-type RE SOs (see Ref. [122] and references therein) as well as of B-type and A-type RE SOs.[115,123,124] It should be noted that powder XRD is not able to discern between the trigonal A-type phase and the hexagonal H-type phase since both yield very similar powder XRD patterns.[125] However, these two phases have sizably different vibrational properties due to the fact that the H-type phase has twice the number of formula units than the A-type phase. As will become evident below, the present RS measurements in $Tb_2O_3$ at HP confirm both the cubic-to-monoclinic (C→B) and the monoclinic-to-trigonal (B→A) PTs in this compound.

Group theory predicts that the cubic C-type structure ($Ia\bar{3}$) should have 120 zone-center vibrational modes: $\Gamma_{120}= 4A_g$ (R) $+ 4E_g$ (R) $+ 14F_g$ (R) $+ 5A_u + 5E_u + 17F_u$ (IR), where $E$ and $F$ modes are double and triple degenerate, respectively ($F$ modes are also denoted as $T$ in some works). The 22 gerade ($g$) modes are Raman-active (R) modes, the 16 $F_u$ modes are infrared (IR)-active modes, the $A_u$ and $E_u$ are silent modes, and one $F_u$ mode corresponds to the acoustic vibrations. As regards the monoclinic B-type structure ($C2/m$), group theory predicts 42 zone-center vibrational modes: $\Gamma_{42}= 14A_g$ (R) $+ 7B_g$ (R) $+ 7A_u$ (IR) $+ 14B_u$ (IR) $+ A_u + 2B_u$. The 21 gerade ($g$) modes are Raman-active (R) modes, while 21 ungerade ($u$) modes are infrared (IR)-active modes, and one $A_u$ and two $B_u$ modes correspond to the acoustic vibrations. Finally, the trigonal A-type structure ($P\bar{3}m1$) should have the following 15 zone-center vibrational modes: $\Gamma_{15}= 2A_{1g}$ (R) $+ 2E_g$ (R) $+ 2A_{2u}$ (IR) $+ 2E_{2u}$ (IR) $+ A_{2u} + E_u$. The 4 $g$ modes are Raman-active, the 4 $u$ modes are IR-active, and one $A_{2u}$ and one $E_u$ modes correspond to the acoustic vibrations. It should be noted that group theory predicts that the hexagonal H-type structure ($P6_3/mmc$) should have 30 zone-center vibrational modes because it has two formula units per unit cell: $\Gamma_{30}= 2A_{1g}$ (R) $+ 2E_{1g}$ (R) $+ 2E_{2g}$ (R) $+ 2A_{2u}$ (IR) $+ 3E_{1u}$ (IR) $+ 3B_{2u} + 3E_{2u} + A_{2u} + E_{1u}$. The 6 $g$ modes are Raman-active, 5 $u$ modes are infrared (IR)-active and 6 $u$ modes are silent, and one $A_{2u}$ and one $E_{1u}$ modes correspond to the acoustic vibrations. Therefore, as will be evident below, the number of observed Raman-active modes may allow one to distinguish between the A (trigonal) and H (hexagonal) structures.

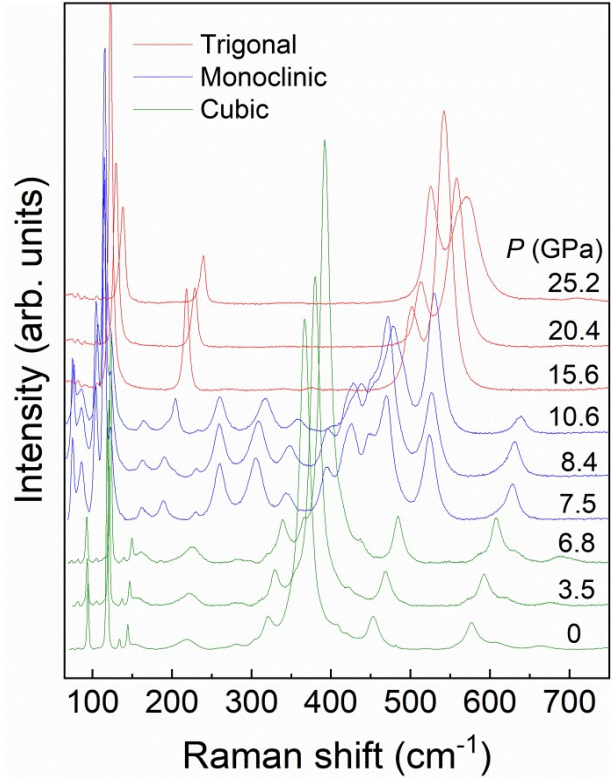

**Figure 10:** Room-temperature Raman spectra of $Tb_2O_3$ at selected pressures up to 25 GPa.

We have recently characterized Raman-active modes of C-type $Tb_2O_3$ at room pressure.[89] In the present work, we have performed unpolarized RS measurements and observed up to 16 out of 22 Raman-active modes of the cubic phase as a function of pressure (see **Fig. 10**). It can be observed that the cubic phase is stable from 0 to 6.8 GPa, and has completely disappeared at 7.5 GPa. Above 6.8 GPa, new Raman peaks emerge in the spectra that have been attributed to the monoclinic B-type phase. In particular, a very weak feature at 185 cm$^{-1}$ is observed at that pressure, despite the new phase is only clear above 7.5 GPa. RS spectra indicate that the monoclinic phase is stable up to 10.6 GPa, in agreement with our XRD results. At 11.5 GPa, other Raman peaks showed up, as evidenced by the appearance of features at 490 and 530 cm$^{-1}$, that have been attributed to the trigonal A-type phase. Finally, the A phase was found to back-transit, on the downstroke, to the B-type phase below 10.4 GPa. The B-type phase was fully recovered at 8.4 GPa and, similarly to our XRD results, it was found to be metastable down to 0 GPa. In order to help in the identification of this metastable phase, the RS spectrum of the B-type phase at room pressure is plotted in **Fig. 11**, together with the theoretical zero-pressure frequencies according to GGA-PBE and GGA-PBEsol DFT calculations. As can be seen in the figure, the PBE functional provides better results in the case of the low-frequency modes (<250 cm$^{-1}$). In contrast, PBEsol seems to be in better agreement above 250 cm$^{-1}$. While PBEsol clearly underestimates the unit-cell volume of both the C-, B- and A-type phases of Tb$_2$O$_3$, the reduced interatomic distances predicted in this case seem to yield increased spring constants and therefore improved results in the case of the high-frequency modes (with mainly O vibrations as shown in Ref. [89]). From now on, we rely on the GGA-PBE vibrational results as this functional better reproduces both the compression behavior of the unit-cell volume and the low-frequency modes (with mainly Tb vibrations).

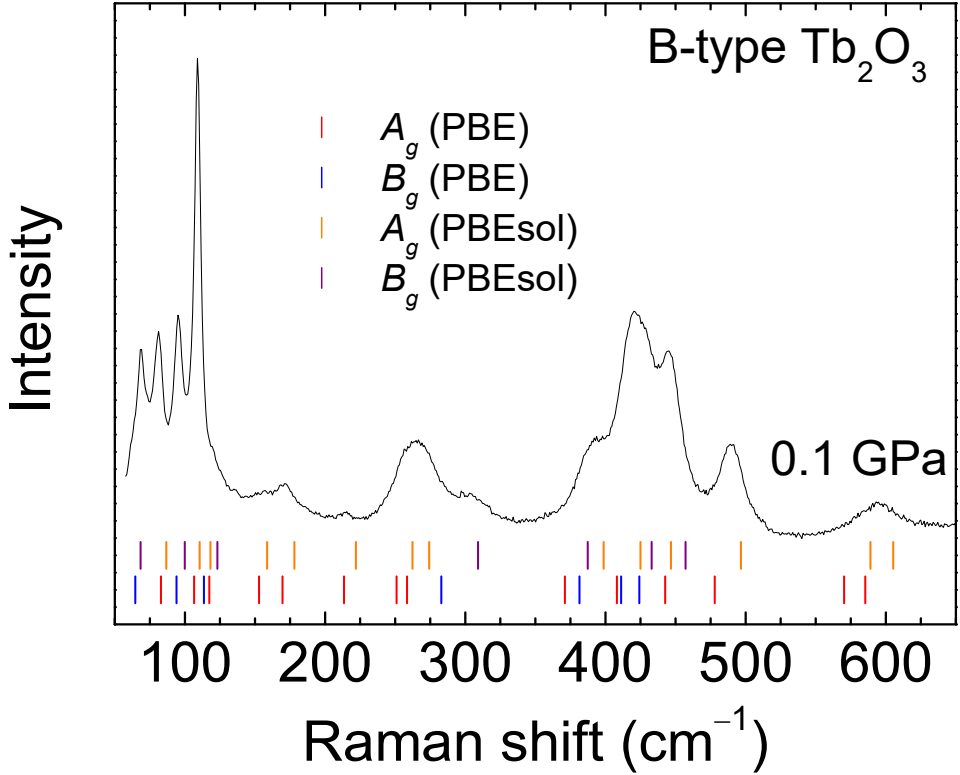

**Figure 11:** Room-temperature Raman spectrum of the recovered monoclinic B-type phase of Tb$_2$O$_3$ at room pressure. Pressure is somewhat larger (~0.1 GPa) than ambient pressure because the DAC was not open to perform this measurement. The vertical ticks indicate calculated frequencies for $A_g$ and $B_g$ modes with DFT-PBE (lower ticks) or DFT-PBESol (upper ticks).

**Figure 12** shows the experimental and theoretical wavenumbers of the Raman-active modes in Tb$_2$O$_3$ as a function of pressure up to 25 GPa. The experimental data correspond to the upstroke measurements. As can be seen in the plot, the pressure dependence of the theoretical wavenumbers of the Raman-active modes is in good agreement with our experimental results and almost all frequencies shift to higher frequencies due to the shortening of bond lengths as a consequence of the volume decrease with increasing pressure. Besides, Raman data confirm not only the two phase transitions observed by HP-XRD but also that the trigonal A-type nature of the HP phase above 12 GPa, as it only displays 4 Raman-active modes.

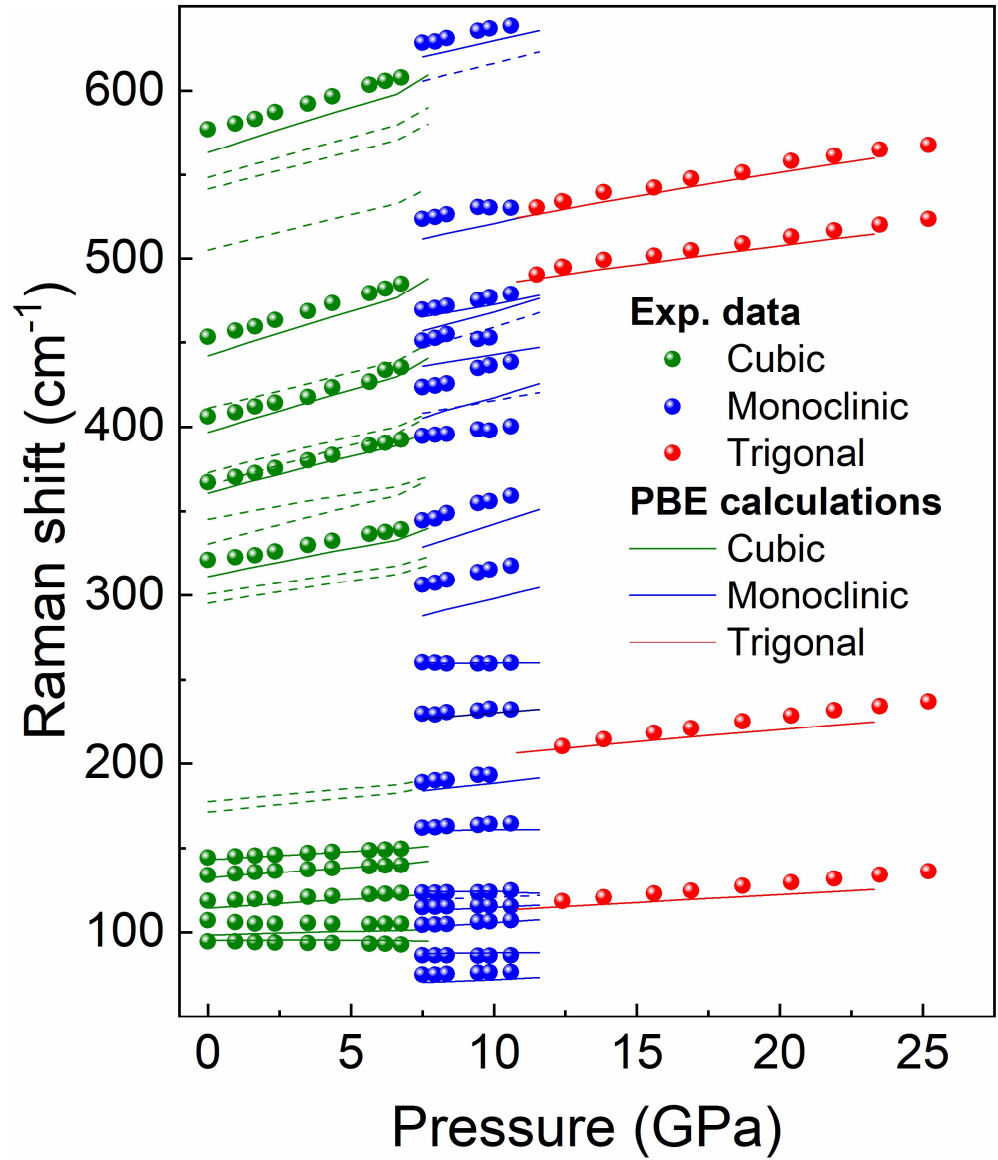

**Figure 12**: Experimental (symbols) and theoretical DFT-PBE (lines) pressure dependence of the Raman-active mode frequencies in cubic (green), monoclinic (blue) and trigonal (red) $Tb_2O_3$ on increasing pressure. Solid (dashed) lines correspond to observed (non-observed) Raman active modes.

We have plotted in **Fig. 13** the pressure dependence of the experimental and theoretical Raman-active mode frequencies of B- and A-type $Tb_2O_3$ on downstroke from 25 GPa down to room pressure, but also including the experimental upstroke values of these two phases shown in **Fig. 12**. These data have allowed us to get the full pressure dependence of the Raman-active modes of the B-type phase. The experimental and theoretical wavenumbers and pressure coefficients of the Raman-active optical phonons of $Tb_2O_3$ in the three phases (C-, B- and A-type) can be found in **Tables 3, 4 and 5**, respectively. As observed in those tables and in **Figs. 11 and 12**, a good agreement is found between our experimental and theoretical lattice-dynamical results. Moreover, the comparison between the calculated and experimental frequencies and pressure coefficients allows us to assign the symmetries of the different Raman peaks from the two HP phases with a high degree of certainty. This is particularly useful, as polarized measurements are not feasible for these phases. The symmetries of the different modes are displayed in the first column of these three tables.

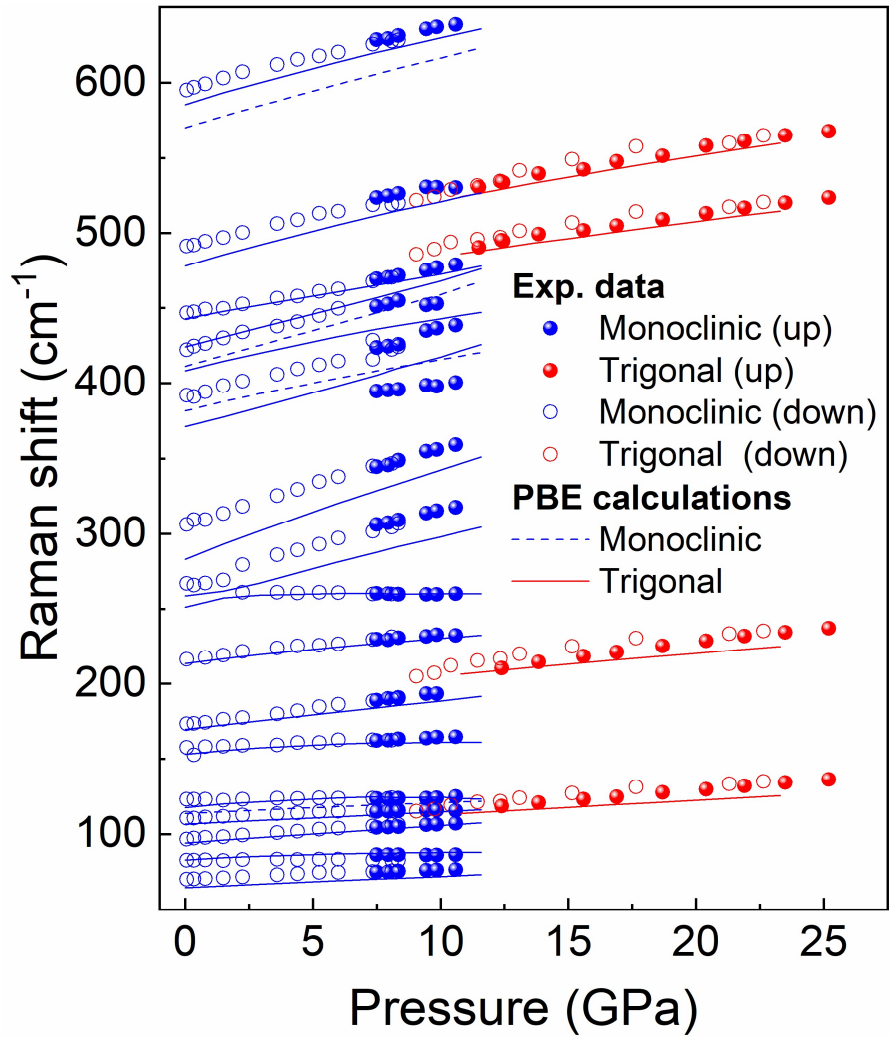

**Figure 13**: Experimental (symbols) and theoretical DFT-PBE (lines) pressure dependence of the Raman-active mode frequencies in monoclinic and trigonal $Tb_2O_3$. Experimental data on increasing pressure (solid symbols) and decreasing pressure (empty symbols) are plotted for comparison. Solid (dashed) lines correspond to observed (non-observed) Raman-active modes.

**Table 3**: Experimental (upstroke) and DFT-PBE theoretical zero-pressure wavenumbers (in cm$^{-1}$) and linear pressure coefficients (in cm$^{-1}$GPa$^{-1}$) for the Raman-active modes of C-type Tb$_2$O$_3$.

| Symmetry | Theoretical | | Experimental | |
|---|---|---|---|---|
| | $\omega_0$ | $d\omega/dP$ | $\omega_0$ | $d\omega/dP$ |
| $F_g^1$ | 95.05 | −0.06 | 94.5 | −0.25 |
| $F_g^2$ | 98.36 | 0.36 | 106.2 | −0.24 |
| $A_g^1$ | 114.29 | 0.96 | 118.6 | 0.68 |
| $F_g^3$ | 132.35 | 1.13 | 134.2 | 0.87 |
| $E_g^1$ | 143.13 | 0.93 | 144.4 | 0.78 |
| $F_g^4$ | 171.36 | 1.70 | | |
| $F_g^5$ | 177.48 | 1.52 | | |
| $F_g^6$ | 295.32 | 2.61 | | |
| $F_g^7$ | 300.68 | 2.57 | | |
| $E_g^2$ | 311.07 | 3.32 | 320 | 2.86 |
| $F_g^8$ | 330.57 | 4.33 | | |
| $A_g^2$ | 345.04 | 2.92 | | |
| $F_g^9$ | 360.67 | 4.26 | 366.7 | 3.86 |
| $E_g^3$ | 365.13 | 4.74 | | |
| $F_g^{10}$ | 372.74 | 4.08 | | |
| $A_g^3$ | 396.77 | 5.08 | 404.7 | 4.33 |
| $F_g^{11}$ | 411.15 | 4.44 | | |
| $F_g^{12}$ | 442.06 | 5.32 | 452.4 | 4.76 |
| $F_g^{13}$ | 505.15 | 4.28 | | |
| $A_g^4$ | 541.45 | 4.51 | | |
| $E_g^4$ | 548.42 | 4.83 | | |
| $F_g^{14}$ | 563.49 | 5.30 | 576.4 | 4.76 |

The pressure dependence of the Raman-active mode frequencies of C-type Tb$_2$O$_3$ shown in **Fig. 12** is similar to that reported for C-type Lu$_2$O$_3$,[16,17] Yb$_2$O$_3$,[19] Tm$_2$O$_3$,[21] Sc$_2$O$_3$,[50] Y$_2$O$_3$,[51] and In$_2$O$_3$.[69] Similarly, the pressure behavior of the Raman spectrum of B-type Tb$_2$O$_3$ shown in **Figs. 11 and 12** is similar to that reported in the literature for B-type Lu$_2$O$_3$,[16,17] Yb$_2$O$_3$,[19] Tm$_2$O$_3$,[21] Y$_2$O$_3$,[51] and Sm$_2$O$_3$.[47,126] Finally, the pressure dependence of Raman-active modes of A-type Tb$_2$O$_3$ shown in **Figs. 11 and 12** is similar to that reported for A-type Sm$_2$O$_3$[47,126] and Nd$_2$O$_3$.[127] The wavenumbers and pressure coefficients for the Raman-active modes in the different phases of Tb$_2$O$_3$ can be compared with those of other RE SOs. In **Tables S3 and S4** in the SI we have compared the experimental and theoretical Raman mode frequencies and pressure coefficients of several C-type SOs, respectively. On the other hand, **Tables S5 and S6** in the SI we have compared the experimental Raman mode frequencies and pressure coefficients of several B-type and A-type SOs, respectively.

As regards the lattice dynamics of C-type SOs, several works have recently addressed the comparison of Raman-mode frequencies in C-type SOs.[89,122] In this work, we will concentrate on the comparison of the pressure coefficients provided in **Tables S3 and S4**. In particular, the large difference in mass of the RE ions and O has allowed us to distinguish two regions: the low-frequency region (below 250 cm$^{-1}$) dominated by pure RE vibrations and the high-frequency region (above 250 cm$^{-1}$) dominated by pure oxygen vibrations. One additional subdivision can be added to the high-frequency region since Raman-active modes in C-type RE SOs show phonon gaps between the different regions. In this way, we can stablish: i) the low-frequency region below 250 cm$^{-1}$; ii) the medium-frequency region between 250 and 550 cm$^{-1}$; and iii) the high-frequency region above 550 cm$^{-1}$. The modes in the medium-frequency region are generally bending RE-O modes, while those in the high-frequency region are generally stretching RE-O modes. In general, the theoretical pressure coefficients of the vibrational modes increase with increasing frequency. This means that the smallest pressure coefficients must be found in the low-frequency region, while the largest pressure coefficients are found in the medium- and high-frequency regions. In particular, the three highest-frequency modes ($A_g^4$, $E_g^4$ and $F_g^{14}$) are always among the Raman modes with the highest pressure coefficients. Other modes with large pressure coefficients are the $E_g^3$, $A_g^3$ and $F_g^{12}$ modes. On the contrary, the two Raman modes with the lowest frequency ($F_g^1$ and $F_g^2$) are always among the Raman modes with the smallest pressure coefficients, which is some cases can be negative, like in C-type Tb$_2$O$_3$.

**Table 4**: Experimental and DFT-PBE theoretical zero-pressure wavenumbers (in cm$^{-1}$) and linear pressure coefficients (in cm$^{-1}$GPa$^{-1}$) for the Raman-active modes of B-type Tb$_2$O$_3$. For the fits of experimental data, the upstroke and downstroke results have been used. The pressure dependence of all wavenumbers has been fitted to a linear polynomial in the low-pressure range (0-2 GPa).

| Symmetry | Theoretical | | Experimental | |
|---|---|---|---|---|
| | $\omega_0$ | $d\omega/dP$ | $\omega_0$ | $d\omega/dP$ |
| $B_g^1$ | 64.58 | 0.73 | 70.3 | 0.7 |
| $A_g^1$ | 82.90 | 0.37 | 82.9 | 0.01 |
| $B_g^2$ | 94.05 | 1.15 | 96.8 | 1.16 |
| $A_g^2$ | 106.55 | 0.82 | 110.8 | 0.64 |
| $A_g^3$ | 113.64 | 0.68 | | |
| $B_g^3$ | 117.37 | 0.39 | 122.9 | 0.13 |
| $A_g^4$ | 152.90 | 0.61 | 156.3 | 0.88 |
| $A_g^5$ | 169.61 | 1.89 | 172.8 | 2.14 |
| $A_g^6$ | 213.53 | 1.57 | 216.5 | 1.8 |
| $A_g^7$ | 250.99 | 0.47 | 261.8 | −0.23 |
| $A_g^8$ | 258.47 | 4.13 | 265.2 | 5.09 |
| $B_g^4$ | 282.94 | 5.77 | 306.6 | 5.12 |
| $B_g^5$ | 371.05 | 4.73 | 368.6 | 1.73 |
| $A_g^9$ | 381.52 | 3.29 | | |
| $A_g^{10}$ | 408.19 | 3.27 | 391.7 | 3.77 |
| $B_g^6$ | 411.25 | 4.86 | | |
| $A_g^{11}$ | 424.18 | 4.48 | 426.2 | 2.43 |
| $B_g^7$ | 442.58 | 3.03 | 446.0 | 2.95 |
| $A_g^{12}$ | 477.94 | 4.13 | 492.0 | 3.59 |
| $A_g^{13}$ | 570.26 | 4.49 | | |
| $A_g^{14}$ | 585.39 | 4.24 | 596.7 | 3.92 |

**Table 5**: Experimental (upstroke) and DFT-PBE theoretical wavenumbers (in cm$^{-1}$) at 11 GPa and linear pressure coefficients (in cm$^{-1}$GPa$^{-1}$) for the Raman-active modes of A-type Tb$_2$O$_3$.

| Symmetry | Theoretical | | Experimental | |
|---|---|---|---|---|
| | $\omega(P = 11$ GPa$)$ | $d\omega/dP$ | $\omega(P = 11$ GPa$)$ | $d\omega/dP$ |
| $E_g^1$ | 113.7 | 0.95 | 116.0 | 1.40 |
| $A_{1g}^1$ | 206.6 | 1.47 | 207.7 | 2.07 |
| $A_{1g}^2$ | 486.6 | 2.28 | 490.3 | 2.32 |
| $E_g^2$ | 524.7 | 2.87 | 529.2 | 2.79 |

The negligible or even negative pressure coefficient of the lowest frequency modes of C-type Tb$_2$O$_3$ is similar to that observed in other C-type sesquioxides (see Tables 3, S3 and S4). This behavior is also observed in many materials with cubic structures or derived from them, like zincblende or wurtzite compounds such as ZnO[128] or AlN,[129] and $AB_2X_4$ chalcogenides.[130,131] It has also been observed in zircon-type $AB$O$_4$ compounds.[132,133] Certainly, this anomalous decrease of the frequency with increasing pressure cannot be explained by an increase of the cation-anion distances with increasing pressure and it could be related to an instability of these structures derived from the cubic lattice that show negative pressure coefficients of vibrational modes at the Brillouin zone edge.[134] In more complex materials (with lower symmetry or with a large number of formula units per primitive cell), these vibrations become Raman or IR-active at the Brillouin zone center (Γ point) due to the folding of the unit cell along certain directions of low symmetry. This seems to be the case of C-type rare-earth sesquioxides and other isostructural sesquioxides.

As regards the lattice dynamics of B- and A-type SOs, there are several works in which a comparison between their Raman-mode frequencies at room pressure is reported.[115,123,135,136] In fact, vibrational modes of both phases are correlated due to the similarity of both B-type and A-type structures as commented in Ref. [115].

However, we do not know any study in which the pressure coefficients of their Raman-active modes have been compared. A characteristic feature of B-type SOs is that their vibrational spectra can be divided into four regions: i) the low-frequency region below 125 cm$^{-1}$; ii) the medium-low-frequency region between 125 and 250 cm$^{-1}$; iii) the medium-high frequency region between 250 and 300 cm$^{-1}$; and iii) the high-frequency region above 300 cm$^{-1}$. Among all these spectral regions, only a small phonon gap seems to be visible between the medium-high- (between 300 and 450 cm$^{-1}$) and high-frequency regions (above 450 cm$^{-1}$). In Tb$_2$O$_3$, such a gap is observed between 450 and 490 cm$^{-1}$. The modes in the whole medium-frequency region are generally bending RE-O modes, while those in the high-frequency region are generally stretching RE-O modes. In any case, we can observe similar trends to those observed for C-type SOs.

In general, the Raman-active mode frequencies of B-type SOs increase with increasing the atomic number; i.e., with decreasing the unit-cell volume. This result means that despite Lu having larger mass than Tb, most modes of B-type $Lu_2O_3$ are observed at higher frequencies than in B-type $Tb_2O_3$ due to the dominance of the lanthanide contraction effect over the cation mass. The largest difference in frequency due to the lanthanide contraction effect is observed in the high-frequency modes. These modes are mainly related to O vibrations so the cation mass is not the dominant factor. On the contrary, both effects get balanced in the low-frequency region, where the vibrational modes have a larger contribution of cations. This explains the similar frequencies of $B_g^1$, $A_g^1$, $B_g^2$, and $A_g^2$ modes along the Sm to Lu series (see **Table S5** in the SI). As for C-type SOs, the theoretical pressure coefficients of the vibrational modes of B-type SOs tend to increase with increasing frequency. This means that the smallest pressure coefficients are observed in the low-frequency region, while the largest pressure coefficients are found in the medium- and high-frequency regions. In particular, the two largest pressure coefficients have been found for the medium-frequency modes $A_g^8$ and $B_g^4$ followed by the three highest-frequency modes ($A_g^{12}$, $A_g^{13}$ and $A_g^{14}$). On the contrary, the two Raman modes with the lowest frequency ($B_g^1$ and $A_g^2$) are always among the Raman modes with the smallest pressure coefficients, with the $B_g^3$, $A_g^4$ and $A_g^7$ modes also being among the ones with the lowest pressure coefficients. In particular, the one for the $A_g^7$ mode is negative in $Tb_2O_3$.

The characteristic phonon modes of A-type SOs have been studied and compared for several SOs.[115,123,126,127,135,136] A distinctive feature of A-type SOs is that they exhibit a phonon gap between the two lowest frequency modes, i.e., the two bending $E_g^1$ and $A_{1g}^1$ modes, and also between the two lowest and the two highest frequency modes, i.e., the stretching $A_{1g}^2$ and $E_g^2$ modes. In fact, the two stretching modes of A-type $Ln_2O_3$ have been shown to correspond to two different $Ln$-O(II) bond distances. As in the case of B-type SOs, we observe similar trends in A-type SOs to those observed in the low-pressure C-type phase. From **Table S7** in the SI we can see that the two stretching Raman-active mode frequencies at 0 GPa of A-type SOs tend to increase with increasing the atomic number; i.e., with decreasing the unit-cell volume. In this way, the two stretching Raman-active modes of A-type $Tb_2O_3$ (extrapolated at 0 GPa) have higher frequency than those of A-type $La_2O_3$ despite La having smaller mass than Tb. The same occurs for $Sm_2O_3$ and $Y_2O_3$. This is in agreement with the correlation found between the zero-pressure frequencies of the two stretching Raman modes and the $c/a$ ratios of the hexagonal unit cell. Note that $c/a$ values around 1.577 (1.617) are expected for A-type $Tb_2O_3$ ($Y_2O_3$) at 0 GPa.[79] This result shows the dominance of the lanthanide contraction effect over the cation mass in the stretching modes. Again, the largest difference in frequency due to the lanthanide contraction effect is observed in the high-frequency modes and the smallest difference is observed in the low-frequency modes, where both effects (cation mass and unit-cell contraction) get balanced (see **Table S7**).

It must be stressed that there is an increasing difference between the frequencies of the two stretching modes, Δω, of A-type $Ln_2O_3$ compounds as the atomic number increases or the unit-cell volume decreases (see **Table S7**). Curiously, A-type $Y_2O_3$ does not seem to follow the trend observed for the rest of A-type $Ln_2O_3$. This has prompted us to think over the linear relationship between the *c/a* ratio and the frequencies of the two stretching modes in A-type RE SOs (see **Fig. 14**). Our extrapolated data of the two stretching mode frequencies for A-type $Tb_2O_3$ and A-type $Y_2O_3$ to room pressure suggest that there is a quadratic trend for A-type RE SOs instead of a linear trend as previously assumed.[115,123,135] Moreover, we can conclude that the relationship seems to be valid for RE SOs, but it does not apply for $Y_2O_3$ nor probably for $Sc_2O_3$.

In a similar way, the comparable frequencies of the two bending modes of A-type RE SOs along the whole RE series also breaks for $Y_2O_3$. However, this disagreement can be solved if we assume that the two low-frequency modes of A-type SOs, whose frequencies are usually below 200 cm$^{-1}$, are related to cation vibrations as in C-type SOs and then strongly influenced by the cation mass. In such a case, we can renormalize the frequencies of $Y_2O_3$ by multiplying the real frequencies of $Y_2O_3$ by $\sqrt{m_Y/m_{Ho}}$ = 0.7342 since the ionic radii of Y and Ho are very similar.[121] After the renormalization, we find that the renormalized frequencies of $Y_2O_3$ fall in the region of other RE SOs (see **Fig. 14**). Therefore, we can conclude that this result indicates that the two low-frequency modes of A-type SOs have a strong influence of cation vibrations. Note that, given the similar atomic mass of *Ln* atoms relative to Y, similar conclusions would have been obtained employing other *Ln* atoms instead of Ho. On the other hand, a similar renormalization can be performed for the high-frequency modes if we multiply by $\sqrt{\mu_Y/\mu_{Ho}}$ = 0.9642, where $\mu_X$ is the reduced mass of *X* and O atoms. In such a case, no drastic change in frequencies is observed for $Y_2O_3$, thus supporting the conclusion that no correlation exists between the high-frequency stretching modes of $Y_2O_3$ and the *c/a* ratio of the A-type structure of RE SOs.

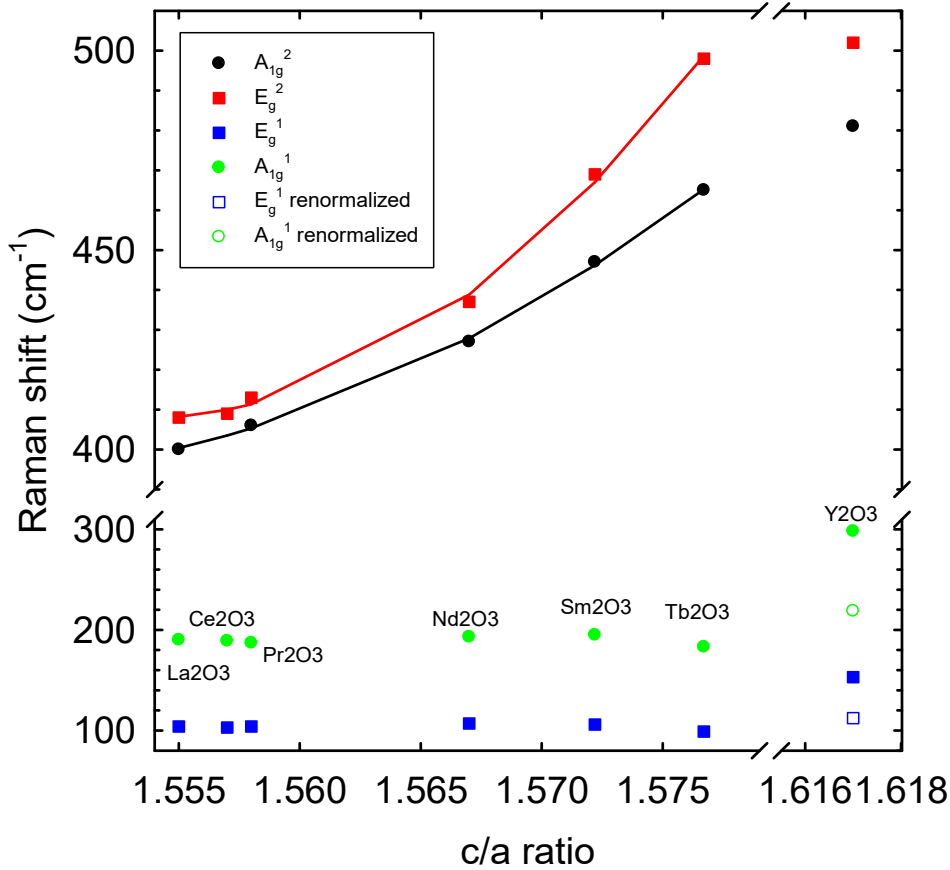

**Figure 14**: Dependence of the experimental Raman-mode frequencies in A-type SOs with the *c/a* ratio. Data have been taken from Table S7 in the Supplementary Material. Renormalized frequency data for $Y_2O_3$ have been calculated as commented in the main text.

The above conclusions regarding the nature of the Raman-active modes of A-type SOs are supported by our *ab initio* calculations of A-type $Tb_2O_3$ at 0 GPa (see **Table S7**). It can be observed that there is a good agreement between our theoretical and extrapolated experimental low-frequency Raman-active mode frequencies at 0 GPa. On the other hand, the high-frequency Raman-active modes show a difference of 2-3% that could be related to the fact that the pressure coefficients of these vibrational modes could tend to decrease as pressure increases thus rendering a not so good extrapolation of experimental modes to 0 GPa. In other words, the experimental pressure coefficients obtained for the high-frequency modes of A-type $Tb_2O_3$ at 11 GPa and listed in **Table S6** are likely smaller than those at 0 GPa. This is consistent with the smaller pressure coefficient of the $E_g^2$ mode in $Tb_2O_3$ at 11 GPa than those found in $Sm_2O_3$, $Nd_2O_3$ and $La_2O_3$ at 0 GPa or close to 0 GPa.

Our calculations have allowed us to represent the atomic vibrations of the Raman-active modes of A-type $Tb_2O_3$ using the J-ICE software.[137] We have clearly observed that the classification of the two low-frequency and the two high-frequency Raman-active modes as pure bending and stretching modes is not correct. First of all, we have to note that there are no pure stretching nor pure bending modes. In all vibrations O(I) atoms are immobile. The low-frequency $E_g^1$ mode corresponds to a mixture of bending and stretching modes where $Ln$ atoms vibrate in the hexagonal $a$-$b$ plane with a small vibration of the O(II) atoms (see **Figure S12a** in the SI). On the other hand, the low-frequency $A_{1g}^1$ mode corresponds to a mixture of a stretching and bending mode where $Ln$ atoms vibrate along the hexagonal $c$ axis against the O(II) atoms (see **Figure S12b** in the SI). All modes of $Tb_2O_3$ below (above) 190 $cm^{-1}$ at 0 GPa (see **Table S8** in the SI) show negligible contribution of the O(II) ($Ln$) atoms. This result means that vibrational modes of the A-type SOs (and likely also for B-type SOs) show similar patterns to those of C-type SOs; i.e. dominance of $Ln$ (O) vibrations below (above) ~200 $cm^{-1}$. As regards the high-frequency modes, the high-frequency $A_{1g}^2$ mode corresponds to a mixture of a stretching and bending mode where $Ln$ atoms vibrate along the hexagonal $c$ axis against the O(II) atoms (see **Figure S13a** in the SI). Finally, the high-frequency $E_g^2$ mode corresponds to a mixture of bending and stretching modes where O(II) atoms vibrate in the hexagonal $a$-$b$ plane with a small vibration of the $Ln$ atoms (see **Figure S13b** in the SI).

As for C- and B-type SOs, in general, the theoretical pressure coefficients of the Raman-active modes of A-type SOs increase with increasing frequency, except for the $A_{1g}^1$ and $A_{1g}^2$ modes of $Nd_2O_3$ and $La_2O_3$ (see **Table S6**). In fact, taking into account the values of the pressure coefficients for $Sm_2O_3$ and $Tb_2O_3$, one can find that the two bending modes are around 106 and 100 $cm^{-1}$ at 0 GPa for the $E_g^1$ mode, respectively, and around 195 and 185 $cm^{-1}$ at 0 GPa for the $A_{1g}^1$ mode, respectively. This result agrees with the previous observation that the two bending modes show almost fixed frequencies around 105 and 190 $cm^{-1}$.[115,123,135] Once again, this trend seems not to be obeyed by A-type $Y_2O_3$, whose extrapolated values at 0 GPa for the two bending modes are around 150 and 300 $cm^{-1}$, respectively. This suggests a different behavior of the bending modes in A-type $Y_2O_3$, and probably also in A-type $Sc_2O_3$, to that in other RE SOs, which points to stronger bending forces in the former. This different behavior, more particularly in the case of $Y_2O_3$, could be related to the slightly lower theoretical bulk moduli predicted for these two materials relative to the rest of RE SOs (see theoretical values in **Fig. 9b)**.

In summary, we have provided RS measurements and lattice dynamics calculations of $Tb_2O_3$ up to 25 GPa and shown that there is a good agreement between experimental and theoretical frequencies for the Raman-active modes of the three phases (C-, B- and A-type) of $Tb_2O_3$. This result confirms the two structural phase transitions already observed by XRD measurements and supported by total-energy *ab initio* calculations. Since B-type $Tb_2O_3$ is metastable at room conditions on downstroke when starting from the B- or the A-type phases, we have provided for the first time the RS spectrum of B-type $Tb_2O_3$ at room pressure and at high pressures that will help future studies to identify this phase by RS measurements instead of resorting to more elaborate XRD measurements. Besides, we have provided the pressure dependence of A-type $Tb_2O_3$ from 11 to 25 GPa. On top of that, we have analyzed in detail the nature of the Raman-active modes of A-type $Tb_2O_3$ in relation to previous works of A-type RE SOs. Finally, we have studied and compared the pressure coefficients of the different Raman-active modes of C-, B- and A-type RE SOs (including $Sc_2O_3$ and $Y_2O_3$) and found a good correlation between the Raman-active modes between the three phases. Notably, we have found that modes below (above) ~200 cm$^{-1}$ correspond mainly to vibrations of *Ln* (O) atoms.

## 5- Conclusions

We have reported a joint experimental and theoretical study of the structural and vibrational properties of cubic terbium sesquioxide at high pressure. Powder x-ray diffraction and Raman scattering measurements up to 25 GPa show that C-type $Tb_2O_3$ undergoes two phase transitions: a first irreversible reconstructive transition to the monoclinic B-type phase at ~7 GPa and a subsequent reversible displacive transition from the monoclinic to the trigonal A-type phase at ~12 GPa. We speculate that the C-B transition is a strong first-order phase transition (volume collapse around 8%) involving a considerable structural reconstruction. Thus, most likely there is a kinetic energy barrier that prevents the system to go back from the B-type structure to the original C-type structure on decreasing pressure (i.e., with decreasing the energy provided to the system). Therefore, the irreversibility of this transition at room temperature can probably be linked to the kinetic-energy barrier effect, as observed in other cases like for instance the zircon-scheelite phase transition in a number of *AB*O$_4$ compounds.[138] On the contrary, the B-A phase transition seems to be a weak first-order phase transition (volume collapse below 2%) and it is a reversible transition at room temperature because it probably implies a much smaller kinetic energy barrier between both phases.

The C→B→A phase transition sequence is in good agreement with those observed in rare-earth sesquioxides crystallizing in the cubic phase as recently reviewed.[15] Additionally, we have provided the experimental equation of state for the three phases that are in rather good agreement with previous works. Furthermore, our *ab initio* theoretical calculations predict phase transition pressures and bulk moduli for the three phases in rather good agreement with experimental results. Moreover, a discussion of the unit-cell volume and bulk moduli in C-type/bixbyite, B- and A-type sesquioxides as a function of the atomic number and of the third power of the cation ionic size has been provided, which serves to check the goodness of both experimental and theoretical results for the three phases in all known sesquioxides. In particular, we have shown that there is a linear trend between the bulk moduli of the three C-, B-, and A-type phases and the cube cation ionic radius.

Raman-active modes of the three phases have been monitored as a function of pressure, and rather good agreement with lattice-dynamics calculations has been found. This agreement has allowed us to confirm the assignment of the experimental phonon modes in the C-type phase as well as to make a tentative assignment of the symmetry of most vibrational modes in the B- and A-type phases. Finally, the Raman-active phonon frequencies and their phonon pressure coefficients for the three phases of $Tb_2O_3$ have been compared with those of other rare-earth sesquioxides. In particular, we have shown that the stretching Raman-mode frequencies of A-type sesquioxides do not scale linearly with pressure as previously assumed and that Raman-mode frequencies of B-type and A-type $Y_2O_3$ at room pressure do not seem to scale with those of other rare-earth sesquioxides and the same seems to be valid for $Sc_2O_3$.

The present work has been devoted to construct a reference framework to understand the high-pressure behavior of the structural and vibrational properties of RE SOs and isostructural compounds. We hope that this study will stimulate further work on different sesquioxides to better constrain their bulk moduli, since some values reported in the literature seem to be quite underestimated or overestimated. We also hope that more work will be carried out on the less well-known A-type phase sesquioxides. In particular, more high-pressure studies should be performed in the following compounds: $Dy_2O_3$, $Ho_2O_3$, $Er_2O_3$, $Tm_2O_3$, $Yb_2O_3$, $Lu_2O_3$ and $Sc_2O_3$. Such studies would complete the sequence of known A-type sesquioxides and would provide a better understanding of the behavior of this phase under compression.

**Acknowledgements**

Authors thank the financial support of Generalitat Valenciana under Project PROMETEO 2018/123-EFIMAT and of the Spanish Ministerio de Economía y Competitividad under Projects MAT2015-71035-R, MAT2016-75586-C4-2/3/4-P, FIS2017-2017-83295-P, as well as through MALTA Consolider Team research network under project RED2018-102612-T. J.A.S. also acknowledges the Ramón y Cajal program for funding support through RYC-2015-17482. A.M. and P.R.-H. acknowledge computing time provided by Red Española de Supercomputación (RES) and the MALTA Consolider Team cluster. HP-XRD experiments were performed at MPSD beamline of Alba Synchrotron (experiment no. 2016071772). We would like to thank Oriol Blázquez (Universitat de Barcelona) for his contribution to the Raman measurements.

**Statement of authorship**

F.J.M. and J.I. prompted the investigation and wrote the manuscript, with contributions of the rest of authors. J.A.S., V.C.G. and F.J.M performed the HP-XRD measurements, assisted by C.P. at ALBA synchrotron. J.A.S. and V.C.G. analyzed the XRD results. J.I. and J.A.S. performed the HP-Raman measurements, and J.I. and R.O. analyzed the data. M. V. and Ph. V. grew the samples used in this work. P.R.H. and A. M. performed all the DFT calculations. Finally, O.G. performed calculations of compressibility tensors and U.R.R.M. contributed to the general discussion of the results

.

**Supporting information**

Whole pattern fittings at different pressures, additional experimental and theoretical structural and vibrational data for C-, A, and B-type $Tb_2O_3$ and isostructural compounds, representation of atomic vibrations for selected phonons of $Tb_2O_3$, and calculation of the experimental and theoretical compressibility tensor of B-type $Tb_2O_3$ at different pressures.

# Supplementary Material of

# Structural and lattice-dynamical properties of Tb$_2$O$_3$ under compression: a comparative study with rare-earth and related sesquioxides

Jordi Ibáñez,[1] Juan Ángel Sans,[2] Vanesa Cuenca-Gotor,[2] Robert Oliva,[3*] Óscar Gomis,[4] Plácida Rodríguez-Hernández,[5] Alfonso Muñoz,[5] Ulises Rodríguez-Mendoza,[5] Matías Velázquez,[6] Philippe Veber,[6,7] Catalin Popescu[8] and Francisco Javier Manjón[2*]

[1]*Institute of Earth Sciences Jaume Almera, MALTA Consolider Team, Consell Superior d'Investigacions Científiques, 08028 Barcelona, Catalonia, Spain*

[2]*Instituto de Diseño para la Fabricación y Producción Automatizada, MALTA Consolider Team, Universitat Politècnica de València, 46022 València, Spain*

[3]*Faculty of Fundamental Problems of Technology, Wroclaw University of Science and Technology, 50-370, Wrocław, Poland*

[4]*Centro de Tecnologías Físicas, MALTA Consolider Team, Universitat Politècnica de València, 46022 València, Spain*

[5]*Departamento de Física, Instituto de Materiales y Nanotecnología, MALTA Consolider Team, Universidad de La Laguna, 38200 San Cristóbal de la Laguna, Tenerife, Spain*

[6]*Univ. Grenoble Alpes, CNRS, Grenoble INP, SIMAP, 38000 Grenoble, France.*

[7]*CNRS, Institut Lumière Matière, Université Claude Bernard Lyon 1, UMR5306, 69622 Villeurbanne, France*

[8]*ALBA-CELLS, MALTA Consolider Team, 08290 Cerdanyola del Vallès (Barcelona), Catalonia, Spain.*

*Corresponding authors: robert.oliva.vidal@pwr.edu.pl; fjmanjon@fis.upv.es

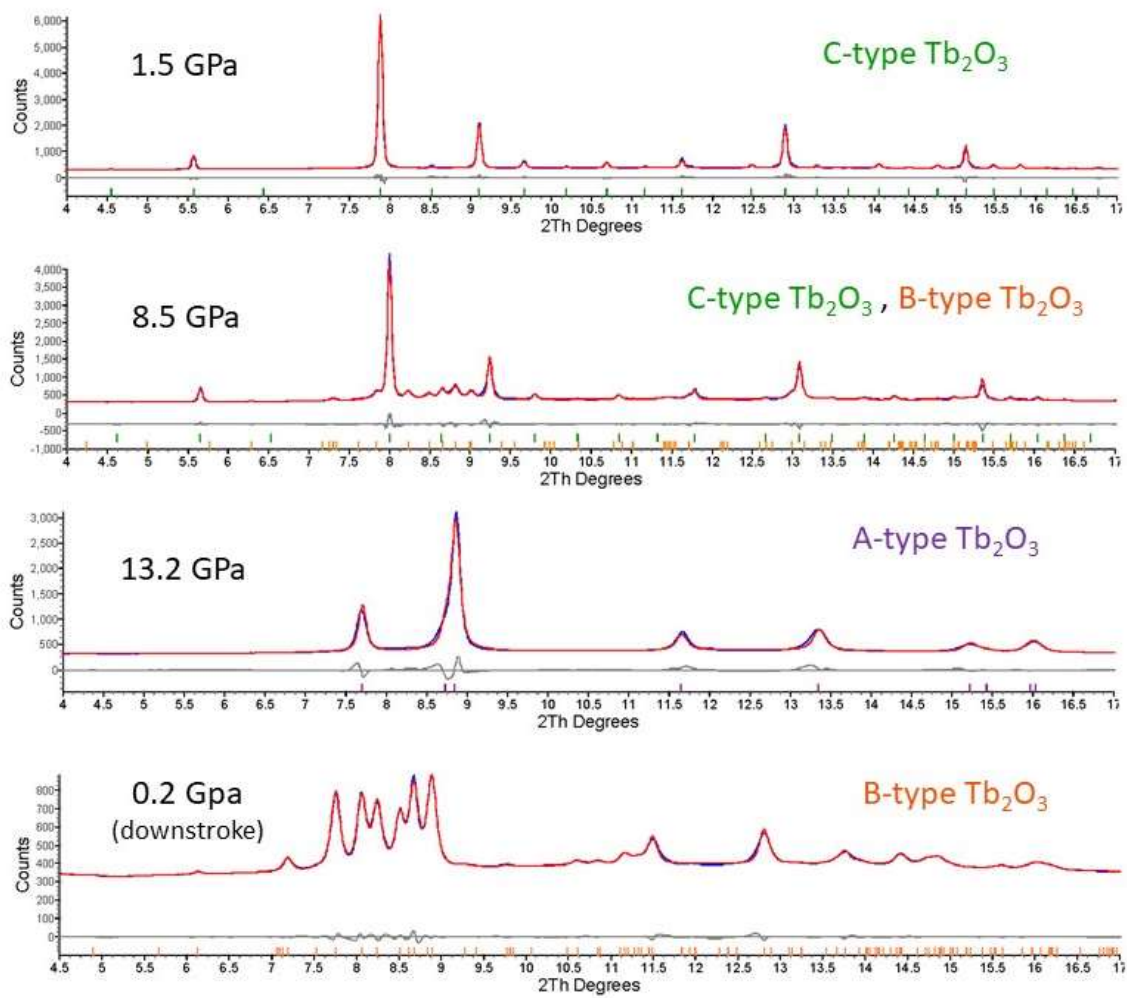

**Figure S1:** Selected examples of calculated and difference profiles obtained from full-pattern matching refinements to the experimental XRD scans. For C-type $Tb_2O_3$, Rietveld refinements were carried out. For the high-pressure polymorphs, the Pawley/Le Bail methods were used.

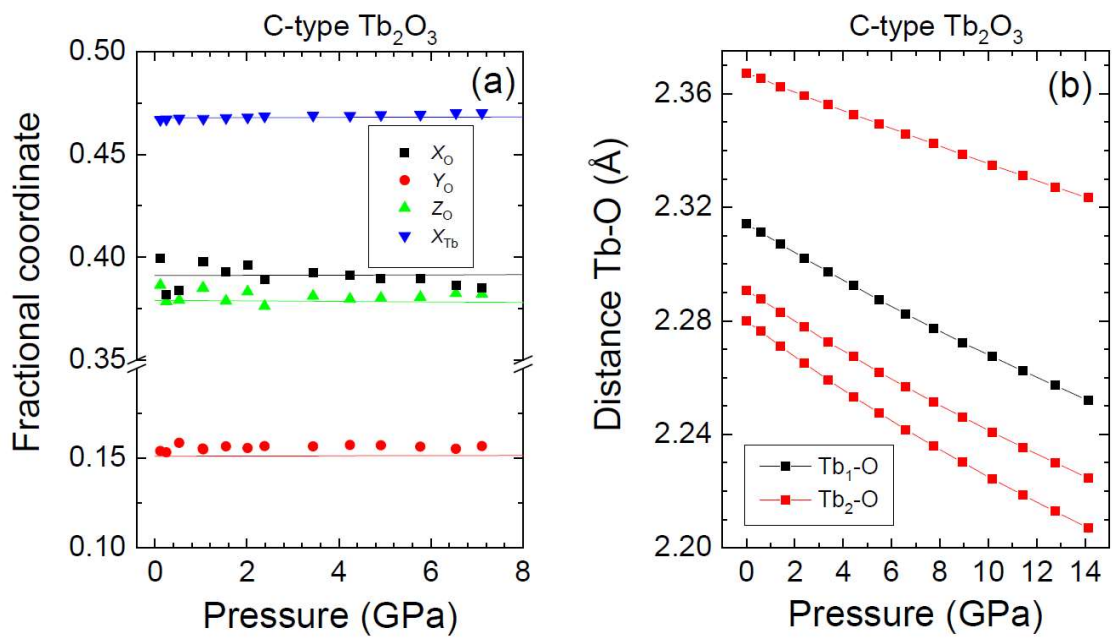

**Figure S2:** (a) Experimental (symbols) and theoretical (lines) pressure dependence of the four free atomic parameters in C-type $Tb_2O_3$. The calculated error for the experimental values is smaller than (of the order of) the size of the symbols for Tb (O) coordinates. (b) Theoretical pressure dependence of the Tb-O distances in C-type $Tb_2O_3$.

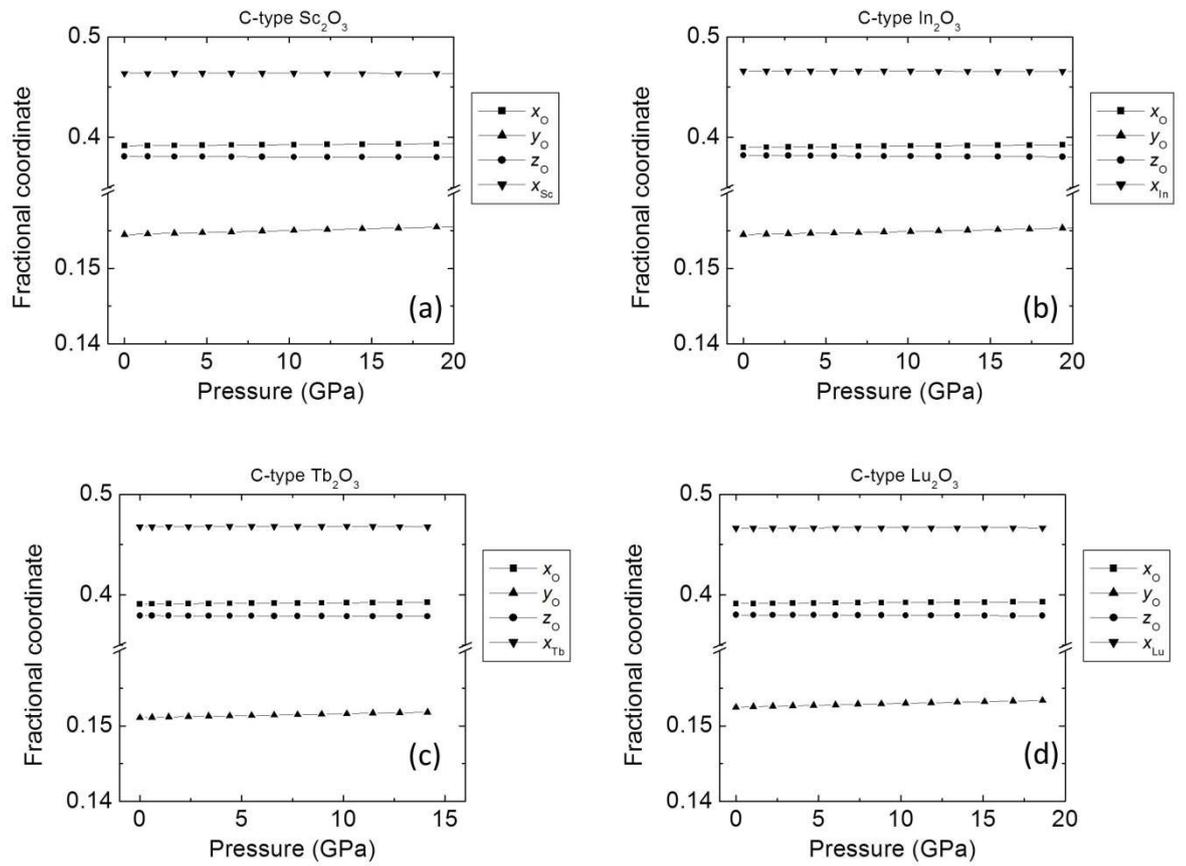

**Figure S3:** Theoretical pressure dependence of the four free atomic parameters in several C-type sesquioxides: (a) $In_2O_3$, (b) $Sc_2O_3$, (c) $Tb_2O_3$, and (d) $Lu_2O_3$.

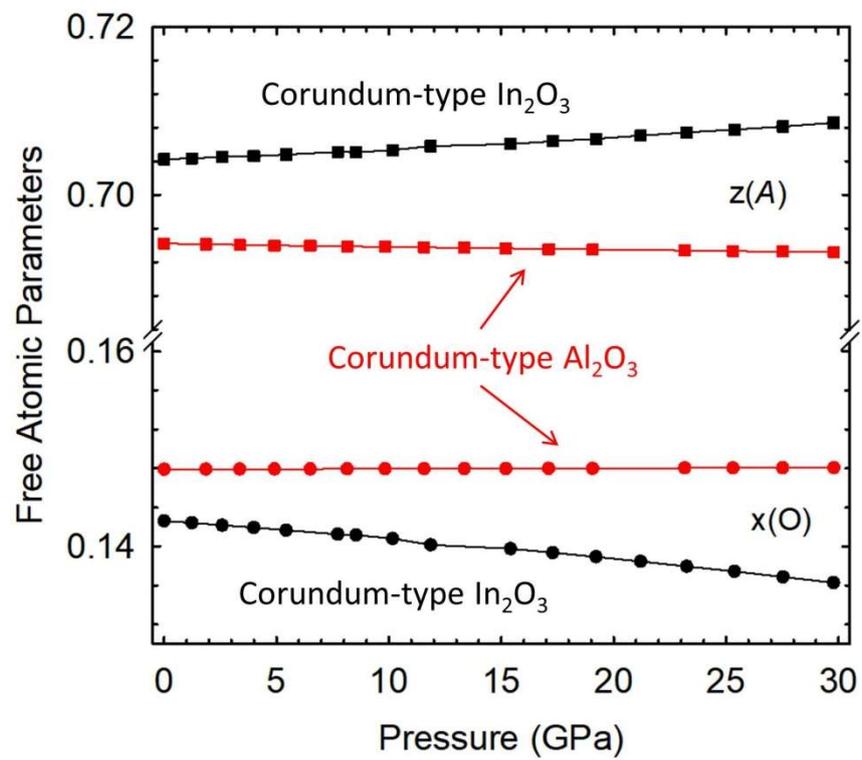

**Figure S4:** Theoretical pressure dependence of the four free atomic parameters in corundum-type $In_2O_3$ and $Al_2O_3$.

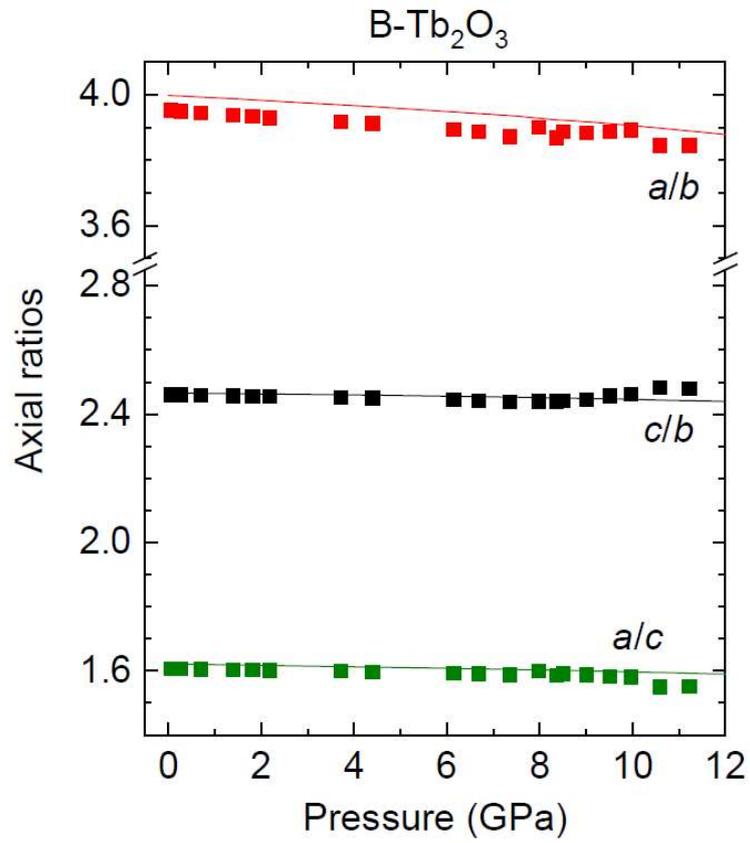

**Figure S5:** Experimental (symbols) and theoretical (lines) pressure dependence of the *c*/*b*, *a*/*b* and *a*/*c* axial ratios in B-type $Tb_2O_3$.

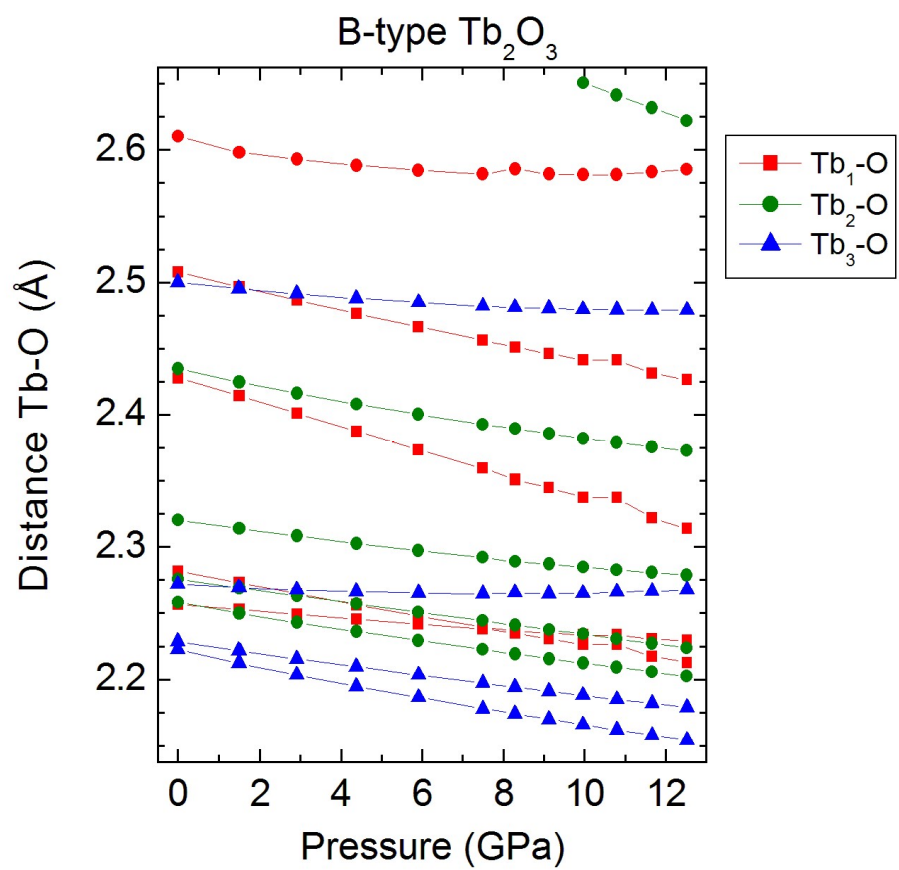

**Figure S6:** Theoretical pressure dependence of the Tb-O distances in B-type $Tb_2O_3$.

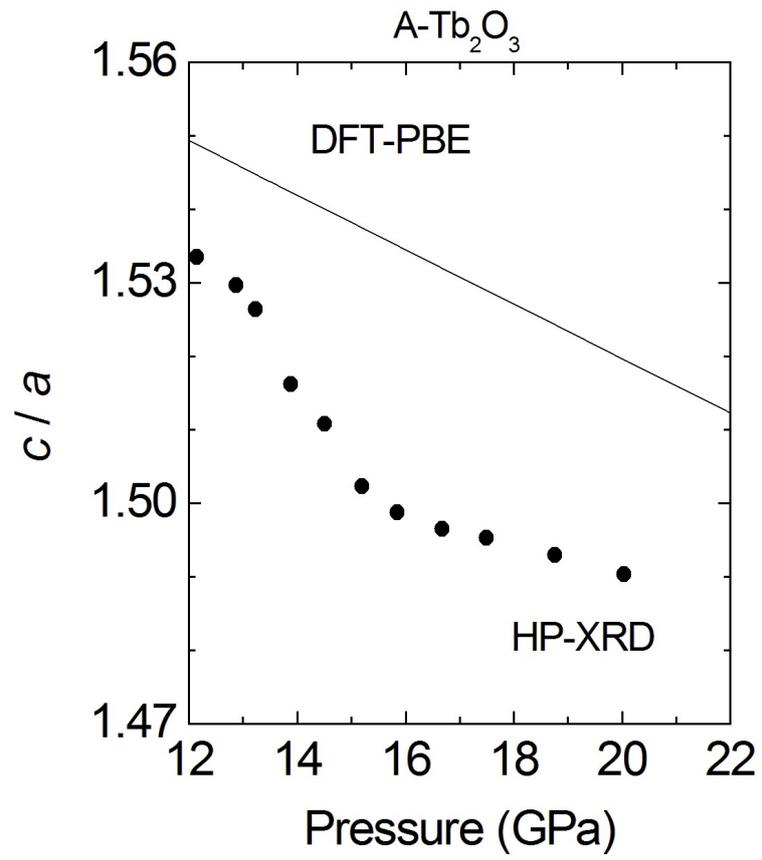

**Figure S7:** Experimental (symbols) and theoretical (lines) pressure dependence of the *c/a* axial ratio in A-type $Tb_2O_3$.

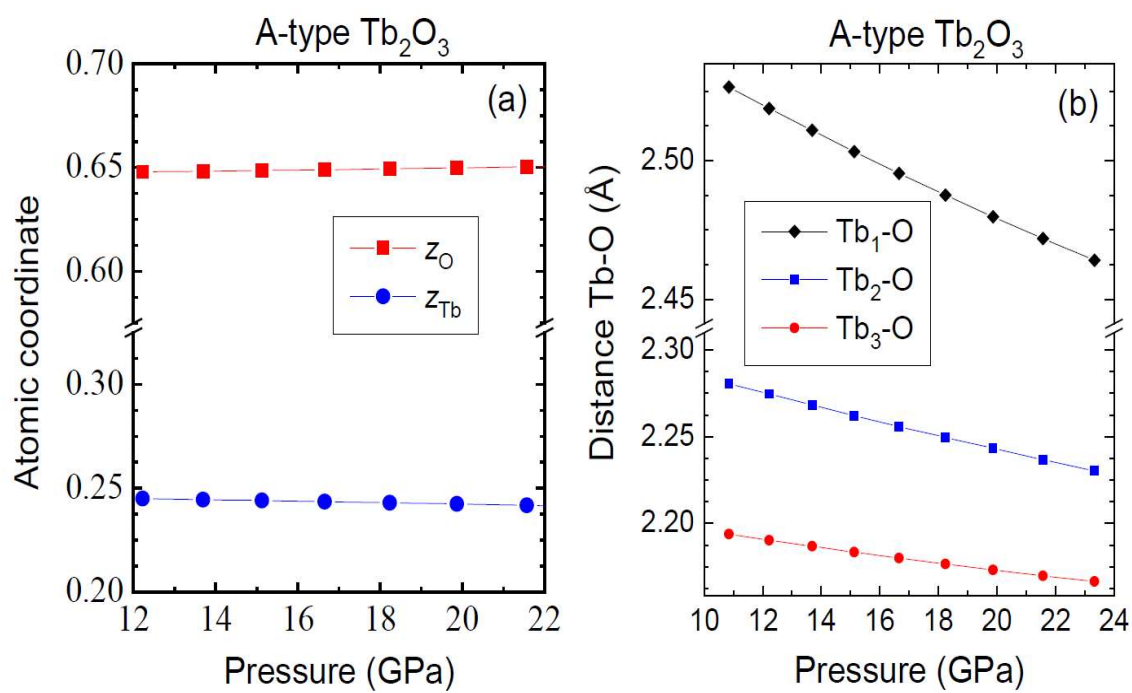

**Figure S8:** (a) Theoretical pressure dependence of the two free atomic parameters of A-type $Tb_2O_3$. (b) Theoretical pressure dependence of the Tb-O distances in A-type $Tb_2O_3$.

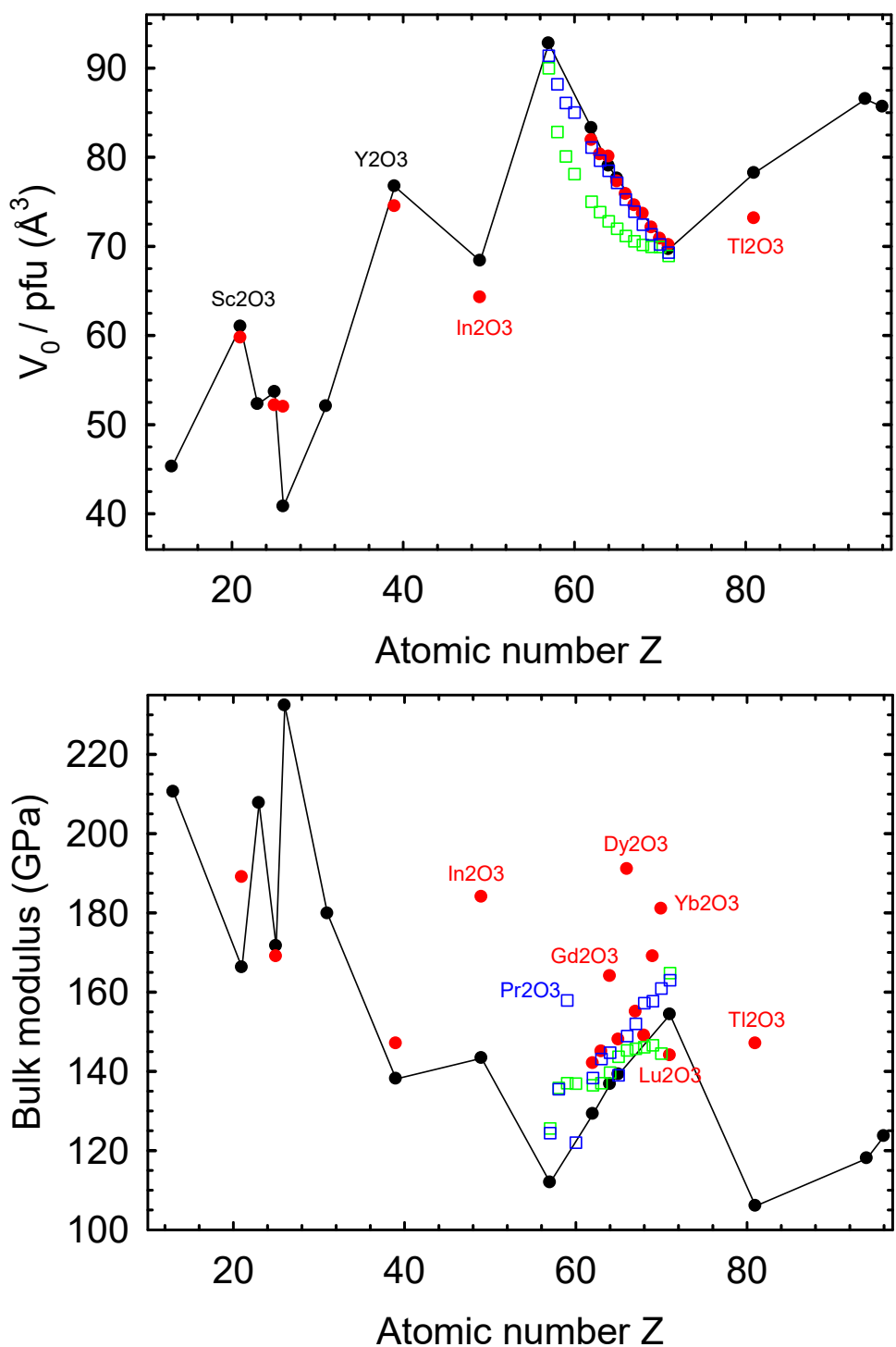

**Figure S9:** Circles represent the experimental (red, from Table S1) and theoretical (black, from Table S2) volume per formula unit (a) and bulk modulus (b) of C-type sesquioxides vs. the atomic number Z. Theoretical data of Ref. 34 with WC-GGA (green open squares) and GGA+U (blue open squares) are also shown for comparison.

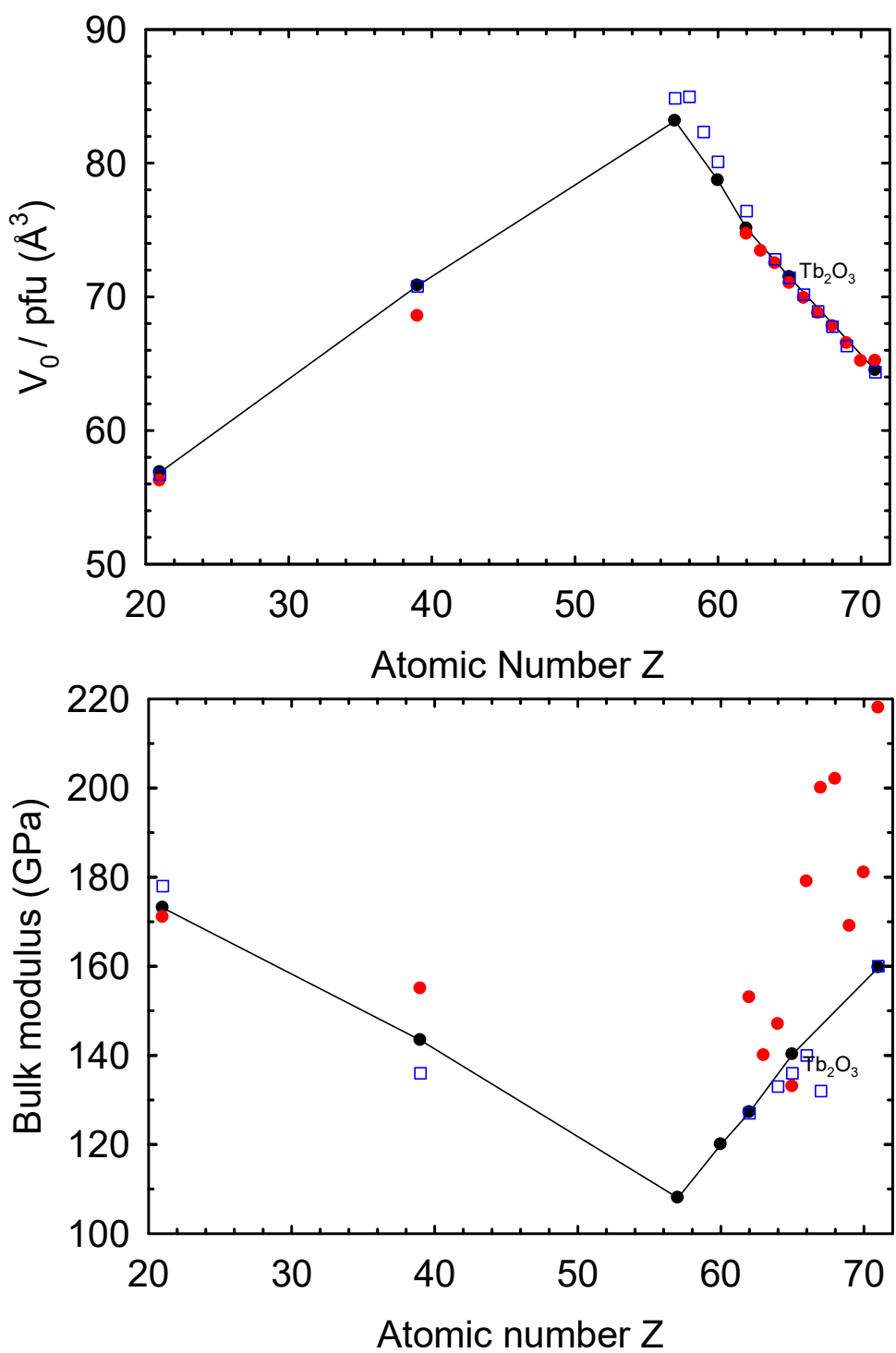

**Figure S10:** Circles represent the experimental (red, from Table S1) and theoretical (black, from Table S2) volume per formula unit (a) and bulk modulus (b) of B-type sesquioxides vs. the atomic number Z. Theoretical data of Ref. 32 (blue open squares) are also shown for comparison.

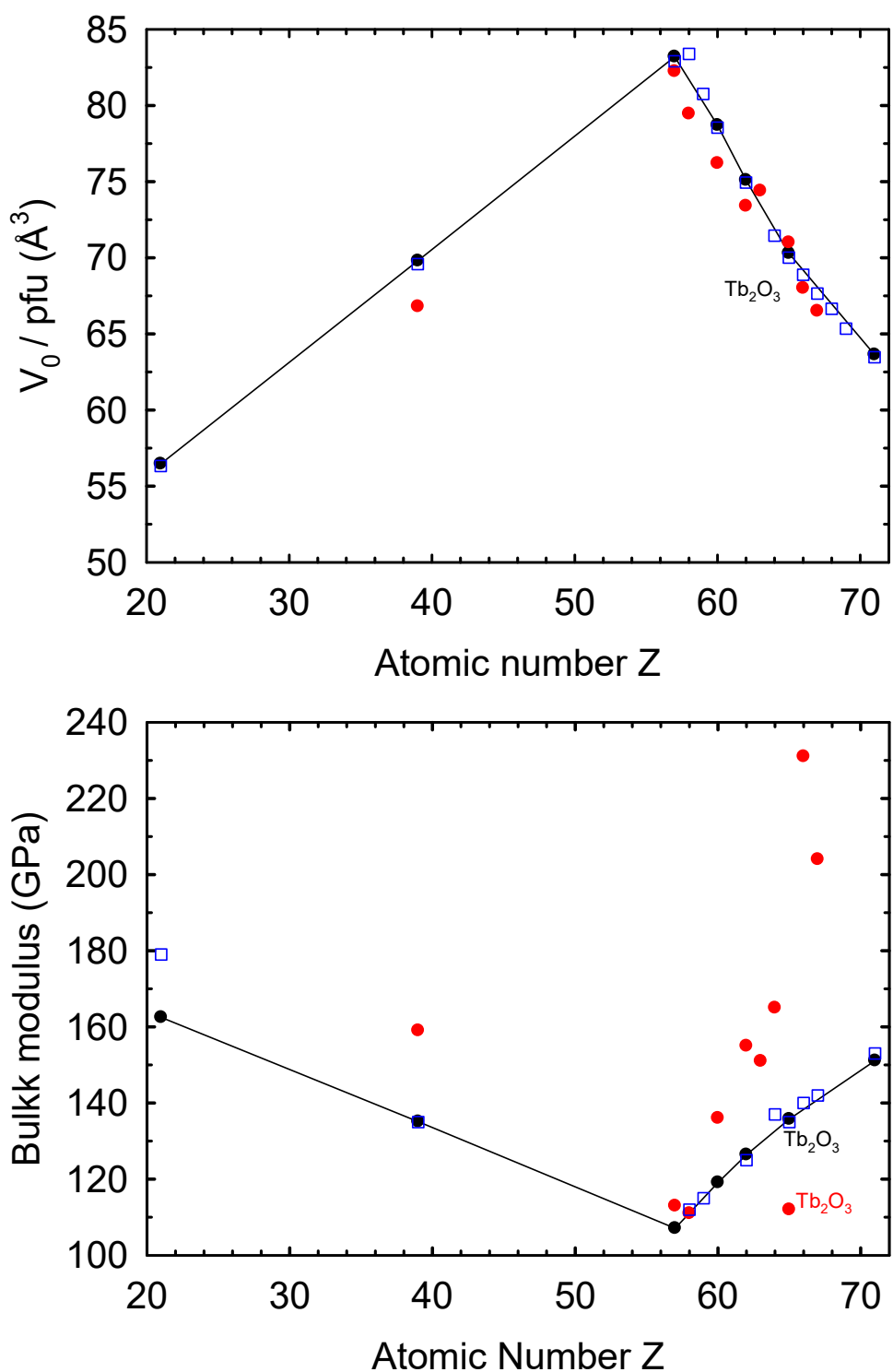

**Figure S11:** Circles represent the experimental (red, from Table S1) and theoretical (black, from Table S2) volume per formula unit (a) and bulk modulus (b) of A-type sesquioxides vs. the atomic number Z. Theoretical data of Ref. 32 (blue open squares) are also shown for comparison.

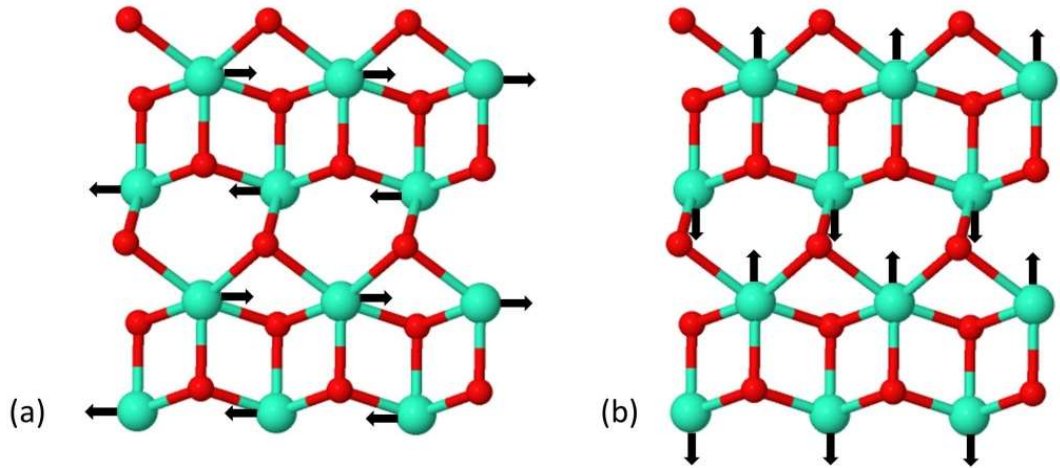

**Figure S12:** Atomic vibrations of the (a) low-frequency $E_g^1$ mode and (b) low-frequency $A_{1g}^1$ modes in A-type $Tb_2O_3$. Big blue atoms are Tb and small red atoms are O. The vertical direction corresponds to the hexagonal c axis and the horizontal direction is perpendicular to the (110) direction of the hexagonal unit cell. Figures have been drawn with the help of J-ICE program [40].

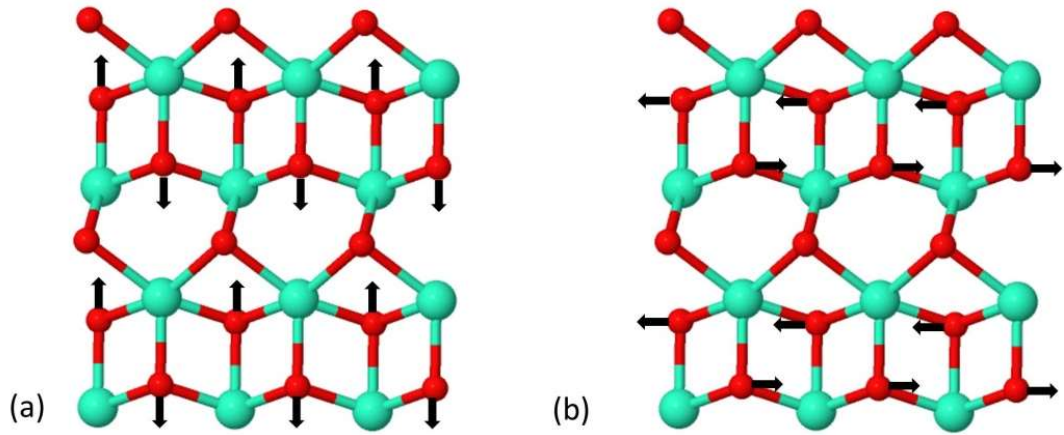

**Figure S13:** Atomic vibrations of the (a) high-frequency $A_{1g}^2$ mode and (b) high-frequency $E_g^2$ modes in A-type $Tb_2O_3$. Big blue atoms are Tb and small red atoms are O. The vertical direction corresponds to the hexagonal c axis and the horizontal direction is perpendicular to the (110) direction of the hexagonal unit cell. Figures have been drawn with the help of J-ICE program [40].

**Table S1**: Experimental data of the equation of state for $Ln_2O_3$ and related sesquioxides having a C-type (or bixbyite) phase, a B-type and an A-type phase. Compounds are ordered on increasing volume per formula unit of the different phases.

| | Phase | $V_0$ (Å$^3$) | $B_0$ (GPa) | $B_0$' | Ref. |
|---|---|---|---|---|---|
| $Fe_2O_3$ | C (exp) | 51.97 | | | [1] |
| $Mn_2O_3$ | C (exp) | 52.2 | 169 | 7.3 | [2] |
| $Sc_2O_3$ | C (exp) | 59.7 | 189 | 4 | [3] |
| | B (exp) | 55.1 | 216 | 5 (fixed) | |
| | C (exp) | 60.8 | 154 | 7 (fixed) | [4] |
| | B (exp) | | 180 | 4 (fixed) | |
| | C (exp) | 59.6 | 198 | 4 (fixed) | [5] |
| | C (exp) | 59.6 | 223 | 1.65 | |
| | B (exp) | 56.2 | 171 | 4 (fixed) | |
| | B (exp) | 57.0 | 141.6 | 4.8 | |
| $In_2O_3$ | C (exp) | 64.9 | 194 | 4.8 (fixed) | [6] |
| | C (exp) | 64.7 | 179 | 5.15 | [7] |
| | C (exp) | 64.3 | 184 | 4 (fixed) | [8] |
| $Lu_2O_3$ | C (exp) | 70.1 | 214 | 9 | [9] |
| | B (exp) | 65.2 | 218 | 2.3 | |
| | C (exp) | 70.1 | 114 | 1.7 | [10] |
| | C (exp) | 70.1 | 144 | 6.7 | [11] |
| $Yb_2O_3$ | C (exp) | 70.8 | 181 | 7.3 | [12] |
| | B (exp) | 65.2 | 181 | 1.3 | |
| $Tm_2O_3$ | C (exp) | 72.1 | 149 | 4.8 | [13] |
| | B (exp) | 66.5 | 169 | 4 (fixed) | |
| $Tl_2O_3$ | C (exp) | 73.1 | 147 | 5 | [14] |
| $Er_2O_3$ | C (exp) | 73.5 | 200 | 8.4 | [15] |
| | B (exp) | 67.8 | 202 | 1.0 | |
| | C (exp) | 73.3 | 136 | 5.9 | [11] |
| | C (exp) | 73.6 | 148.8 | 4.02 | [16] |
| $Y_2O_3$ | C (exp) | 74.4 | 146 | 5.5 | [11] |
| | C (exp) | 74.5 | 147 | 4 (fixed) | [17] |
| | B (exp) | 68.6 | 155 | 4 (fixed) | |
| | A (exp) | 66.8 | 159 | 4 (fixed) | |
| | B (exp) | 69.0 | 159 | 4 (fixed) | [18] |
| | A (exp) | 67.8 | 156 | 4 (fixed) | |

|  | Phase | $V_0$ (Å$^3$) | $B_0$ (GPa) | $B_0$' | Ref. |
|---|---|---|---|---|---|
| Ho$_2$O$_3$ | C (exp) | 74.6 | 206 | 4.8 | [19] |
|  | B (exp) | 68.8 | 200 | 2.1 |  |
|  | A (exp) | 66.5 | 204 | 3.8 |  |
|  | C (exp) |  | 155 | 4 (fixed) | [20] |
|  | A (exp) |  | 249 | 4 (fixed) |  |
| Dy$_2$O$_3$ | C (exp) | 75.9 | 191 | 2.8 | [21] |
|  | B (exp) | 69.9 | 179 | 4.2 |  |
|  | A (exp) | 68.0 | 231 | 3.5 |  |
| Tb$_2$O$_3$ | C (exp) | 77.3 | 148 | 2.1(4) | This work |
|  | B (exp) | 71.0 | 133 | 4 (fixed) |  |
|  | A (exp) | 71.0 | 112 | 4 (fixed) |  |
| Gd$_2$O$_3$ | C (exp) |  | 188 | 4 (fixed) | [22] |
|  | A (exp) |  | 160 | 4 (fixed) |  |
|  | C (exp) |  | 134 | 4 (fixed) | [23] |
|  | B (exp) | 72.5 | 147 | 4 (fixed) |  |
|  | A (exp) |  | 174 | 4 (fixed) |  |
|  | C (exp) | 79.0 | 125 | 4.7 | [11] |
|  | C (exp) | 80.1 | 164 | 4 | [24] |
| Eu$_2$O$_3$ | C (exp) | 80.3 | 145 | 4 | [25] |
|  | A (exp) | 74.4 | 151 | 4 |  |
|  | C (exp) | 80.0 | 115 | 5.9 | [11] |
|  | C (exp) |  | 140 | 4 (fixed) | [20] |
|  | A (exp) |  | 155 | 4 (fixed) |  |
|  | B (exp) | 73.4 | 140 | 4 (fixed) | [26] |
| Sm$_2$O$_3$ | C (exp) | 81.9 | 142 | 4 | [27] |
|  | A (exp) | 72.1 | 224 | 1.5 |  |
|  | C (exp) | 81.7 | 116 | 4 (fixed) | [11] |
|  | A (exp) | 73.0 | 130 | 6.9 |  |
|  | C (exp) | 81.6 | 149 | 4 (fixed) | [28] |
|  | B (exp) | 74.7 | 153 | 4 (fixed) |  |
|  | A (exp) | 73.4 | 155 | 4 (fixed) |  |
| Nd$_2$O$_3$ | A (exp) |  | 136 | 4 (fixed) | [29] |
|  | A (exp) | 76.2 | 142 | 4 (fixed) | [30] |
| Ce$_2$O$_3$ | A (exp) | 79.5 | 111 | 4.7 | [31] |
| La$_2$O$_3$ | A (exp) | 82.2 | 113 | 6.0 | [11] |

**Table S2**: Theoretical data of the equation of state for some $Ln_2O_3$ and related sesquioxides having a C-type (or bixbyite) phase, a B-type and an A-type phase. Compounds are ordered on increasing volume per formula unit of the different phases.

|  | Phase | $V_0$ (Å³) | $B_0$ (GPa) | $B_0$' | Ref. |
|---|---|---|---|---|---|
| $Al_2O_3$ | C (the) | 45.3 | 210.5 | 4.1 | This work |
| $Ga_2O_3$ | C (the) | 52.1 | 179.8 | 4.4 | This work |
| $V_2O_3$ | C (the) | 52.3 | 207.7 | 4.2 | This work |
| $Mn_2O_3$ | C (the) | 53.7 | 171.6 | 4.8 | This work |
| $Sc_2O_3$ | C (the) | 61.0 | 166.2 | 4.2 | This work |
|  | B (the) | 56.9 | 173.1 | 3.9 | This work |
| $In_2O_3$ | C (the) | 68.4 | 143.3 | 4.7 | This work |
| $Lu_2O_3$ | C (the) | 69.7 | 154.3 | 4.3 | This work |
|  | B (the) | 64.5 | 159.7 | 3.6 | This work |
| $Tl_2O_3$ | C (the) | 78.2 | 106 | 5.1 | This work |
| $Y_2O_3$ | C (the) | 76.7 | 138.1 | 4.2 | This work |
|  | B (the) | 70.8 | 143.4 | 3.1 | This work |
|  | A (the) | 69.8 | 135.1 | 3.8 | This work |
| $Tb_2O_3$ | C (the) | 77.5 | 139.2 | 4.3 | This work |
|  | C (the) | 77.3 | 145.2 |  | [31] |
|  | B (the) | 71.5 | 140.2 | 3.3 | This work |
|  | B (the) | 71.4 | 137 | 3.5 | [32] |
|  | A (the) | 70.3 | 135.8 | 3.9 | This work |
|  | A (the) | 70.0 | 136 | 3.5 | [32] |
| $Gd_2O_3$ | C (the) | 79.0 | 136.7 | 4.4 | This work |
| $Sm_2O_3$ | C (the) | 83.2 | 129.2 | 4.5 | This work |
|  | B (the) | 75.1 | 127.2 | 3.8 | This work |
|  | A (the) | 75.1 | 126.4 | 3.8 | This work |
| $Nd_2O_3$ | A (the) | 78.7 | 119.0 | 3.9 | This work |
| $La_2O_3$ | C (the) | 92.8 | 111.9 | 4.3 | This work |
|  | A (the) | 83.2 | 107.1 | 3.9 | This work |

**Table S3**: Experimental zero-pressure wavenumbers (in cm$^{-1}$) and linear pressure coefficients (in cm$^{-1}$GPa$^{-1}$) for the Raman-active modes in several C-type SOs.

| | In$_2$O$_3$[a] | | Lu$_2$O$_3$[b] | | Tm$_2$O$_3$[c] | | Y$_2$O$_3$[d] | | Tb$_2$O$_3$[e] | |
|---|---|---|---|---|---|---|---|---|---|---|
| Sym | $\omega_0$ | $d\omega/dP$ | $\omega_0$ | $d\omega/dP$ | $\omega_0$ | $d\omega/dP$ | $\omega_0$ | $d\omega/dP$ | $\omega_0$ | $d\omega/dP$ |
| $F_g^1$ | 108 | 0.07 | 97 | 0.09 | 99 | 0.01 | | | 95 | −0.25 |
| $F_g^2$ | 118 | 0.4 | | | | | 128 | -0.02 | 106 | −0.24 |
| $A_g^1$ | 131 | 1.0 | 119 | 0.54 | 120 | 0.5 | | | 119 | 0.68 |
| $F_g^3$ | 152 | 1.4 | 136 | 0.80 | | | | | 134 | 0.87 |
| $E_g^1$ | 169 | 0.8 | 146 | 0.77 | | | 163 | 0.67 | 144 | 0.78 |
| $F_g^4$ | 205 | 1.3 | | | | | | | | |
| $F_g^5$ | 211 | 3.0 | | | | | | | | |
| $F_g^6$ | 306 | 2.4 | | | | | | | | |
| $F_g^7$ | | | 332 | 2.51 | | | 317 | 2.24 | | |
| $E_g^2$ | | | | | 337 | 2.66 | 330 | 2.77 | 320 | 2.86 |
| $F_g^8$ | 365 | 4.3 | | | | | | | | |
| $A_g^2$ | 306 | 4.0 | 350 | 2.61 | | | | | | |
| $F_g^9$ | | | | | 383 | 3.66 | 377 | 3.60 | 367 | 3.86 |
| $E_g^3$ | 396 | 3.3 | | | | | | | | |
| $F_g^{10}$ | | | 395 | 3.02 | | | | | | |
| $A_g^3$ | 495 | 3.7 | | | 425 | 4.32 | 426 | 5.58 | 405 | 4.33 |
| $F_g^{11}$ | 467 | | 457 | 3.50 | | | | | | |
| $F_g^{12}$ | 513 | | 503 | 4.46 | 484 | 4.65 | 469 | 4.63 | 452 | 4.76 |
| $F_g^{13}$ | | | | | | | | | | |
| $A_g^4$ | | | | | | | | | | |
| $E_g^4$ | 590 | 5.2 | 598 | 2.06 | | | | | | |
| $F_g^{14}$ | 628 | 6.0 | 618 | 4.21 | 603 | 4.51 | 594 | 4.51 | 576 | 4.76 |

[a]Ref. 8, [b]Ref. 9, [c]Ref. 13, [d]Ref. 35, [e]This work.

**Table S4**: Theoretical zero-pressure wavenumbers (in cm$^{-1}$) and linear pressure coefficients (in cm$^{-1}$GPa$^{-1}$) for the Raman-active modes in several C-type SOs.

| | $In_2O_3$[a] | | $Yb_2O_3$[b] | | $Dy_2O_3$[c] | | $Tb_2O_3$[c] | | $Gd_2O_3$[c] | |
|---|---|---|---|---|---|---|---|---|---|---|
| Sym | $\omega_0$ | $d\omega/dP$ | $\omega_0$ | $d\omega/dP$ | $\omega_0$ | $d\omega/dP$ | $\omega_0$ | $d\omega/dP$ | $\omega_0$ | $d\omega/dP$ |
| $F_g^1$ | 106 | 0.01 | | | 95.02 | 0.12 | 95.05 | −0.06 | 94.60 | 0.13 |
| $F_g^2$ | 114 | 0.30 | | | 98.40 | 0.46 | 98.36 | 0.36 | 97.85 | 0.48 |
| $A_g^1$ | 128 | 0.80 | 114 | 0.97 | 114.10 | 1.06 | 114.29 | 0.96 | 114.00 | 1.18 |
| $F_g^3$ | 148 | 1.20 | | | 131.80 | 1.25 | 132.35 | 1.13 | 132.30 | 1.36 |
| $E_g^1$ | 165 | 1.10 | | | 142.70 | 0.96 | 143.13 | 0.93 | 142.90 | 1.04 |
| $F_g^4$ | 204 | 1.50 | | | 170.90 | 1.70 | 171.36 | 1.70 | 171.10 | 1.89 |
| $F_g^5$ | 211 | 3.00 | | | 176.90 | 1.57 | 177.48 | 1.52 | 177.30 | 1.73 |
| $F_g^6$ | 302 | 2.20 | | | 299.30 | 2.48 | 295.32 | 2.61 | 292.20 | 2.71 |
| $F_g^7$ | 312 | 2.60 | | | 305.00 | 2.41 | 300.68 | 2.57 | 297.10 | 2.63 |
| $E_g^2$ | 308 | 3.10 | 332 | 3.30 | 315.10 | 3.23 | 311.07 | 3.32 | 307.70 | 3.49 |
| $F_g^8$ | 356 | 3.90 | | | 334.00 | 4.34 | 330.57 | 4.33 | 327.60 | 4.67 |
| $A_g^2$ | 302 | 3.10 | | | 348.40 | 3.15 | 345.04 | 2.92 | 342.20 | 3.18 |
| $F_g^9$ | 379 | 3.90 | 379 | 4.17 | 363.80 | 4.24 | 360.67 | 4.26 | 357.80 | 4.62 |
| $E_g^3$ | 385 | 3.60 | | | 368.80 | 4.60 | 365.13 | 4.74 | 362.20 | 5.00 |
| $F_g^{10}$ | 438 | 3.00 | | | 377.70 | 4.29 | 372.74 | 4.08 | 368.30 | 4.64 |
| $A_g^3$ | 476 | 3.40 | 430 | 4.35 | 402.70 | 4.75 | 396.77 | 5.08 | 391.30 | 5.37 |
| $F_g^{11}$ | 447 | 4.30 | | | 416.30 | 3.77 | 411.15 | 4.44 | 406.60 | 4.10 |
| $F_g^{12}$ | 499 | 5.00 | 475 | 5.10 | 448.30 | 5.11 | 442.06 | 5.32 | 436.60 | 5.52 |
| $F_g^{13}$ | 520 | 4.70 | | | 509.70 | 4.00 | 505.15 | 4.28 | 501.30 | 4.28 |
| $A_g^4$ | 576 | 5.40 | | | 547.70 | 4.27 | 541.45 | 4.51 | 536.40 | 4.55 |
| $E_g^4$ | 565 | 5.20 | | | 544.30 | 4.65 | 548.42 | 4.83 | 543.50 | 5.00 |
| $F_g^{14}$ | 600 | 5.40 | 595 | 5.16 | 570.10 | 5.17 | 563.49 | 5.30 | 558.00 | 5.60 |

[a] Ref. 8, [b] Ref. 36, [c] This work

**Table S5**: Experimental zero-pressure wavenumbers (in cm$^{-1}$) and linear pressure coefficients (in cm$^{-1}$GPa$^{-1}$) for the Raman-active modes in several B-type $Ln_2O_3$ compounds.

| Sym. | Lu$_2$O$_3$[a] $\omega_0$ | $d\omega/dP$ | Tm$_2$O$_3$[b] $\omega_0$ | $d\omega/dP$ | Tb$_2$O$_3$[c] $\omega_0$ | $d\omega/dP$ | Sm$_2$O$_3$[d] $\omega_0$ | $d\omega/dP$ |
|---|---|---|---|---|---|---|---|---|
| $B_g^1$ |  |  | 69 | 0.66 | 70.3 | 0.70 |  |  |
| $A_g^1$ |  |  | 84 | 0.78 | 82.9 | 0.01 | 80 | 0.83 |
| $B_g^2$ | 99 | 0.58 | 98 | 1.06 | 96.8 | 1.16 | 95 | 1.62 |
| $A_g^2$ | 113 | 0.53 | 113 | 0.76 | 110.8 | 0.64 | 108 | 1.12 |
| $A_g^3$ |  |  | 128 | 0.68 |  |  |  |  |
| $B_g^3$ | 128 | 0.29 |  |  | 122.9 | 0.13 | 118 | 1.15 |
| $A_g^4$ | 167 | 0.73 | 155 | 1.83 | 156.3 | 0.88 | 144 | 1.67 |
| $A_g^5$ | 174 | 1.56 | 171 | 1.94 | 172.8 | 2.14 | 175 | 3.03 |
| $A_g^6$ |  |  |  |  | 216.5 | 1.8 | 219 | 2.32 |
| $A_g^7$ | 285 | 0.81 |  |  | 261.8 | −0.23 | 244 | 1.42 |
| $A_g^8$ | 313 | 2.07 | 285 | 4.36 | 265.2 | 5.09 | 254 | 6.19 |
| $B_g^4$ |  |  | 321 | 5.38 | 306.6 | 5.12 | 300 | 4.30 |
| $B_g^5$ | 351 | 2.40 |  |  | 368.6 | 1.73 |  |  |
| $A_g^9$ |  |  |  |  |  |  | 345 | 5.30 |
| $A_g^{10}$ |  |  |  |  | 391.7 | 3.77 |  |  |
| $B_g^6$ | 427 | 2.19 | 415 | 3.42 |  |  |  |  |
| $A_g^{11}$ | 458 | 2.53 | 451 | 3.61 | 426.2 | 2.43 | 421 | 3.3 |
| $B_g^7$ | 464 | 2.41 | 479 | 2.71 | 446.0 | 2.95 |  |  |
| $A_g^{12}$ | 540 | 2.36 | 524 | 2.30 | 492.0 | 3.59 | 462 | 4.00 |
| $A_g^{13}$ |  |  | 611 | 3.71 |  |  | 558 | 5.10 |
| $A_g^{14}$ | 640 | 2.01 | 628 | 3.31 | 596.7 | 3.92 | 570 | 4.80 |

[a] Ref. 9, [b] Ref. 13, [c] This work, [d] Ref. 21.

**Table S6**: Experimental wavenumbers (in cm$^{-1}$) and linear pressure coefficients (in cm$^{-1}$GPa$^{-1}$) for the Raman-active phonon modes in several A-type $Ln_2O_3$ compounds. Note that the frequencies correspond to different pressures.

| Sym. | Y$_2$O$_3$[a] $\omega$ | $d\omega/dP$ | Tb$_2$O$_3$[b] $\omega$ | $d\omega/dP$ | Sm$_2$O$_3$[c] $\omega$ | $d\omega/dP$ | Nd$_2$O$_3$[e] $\omega$ | $d\omega/dP$ | La$_2$O$_3$[f] $\omega$ | $d\omega/dP$ |
|---|---|---|---|---|---|---|---|---|---|---|
| $E_g^1$ | 174 | 0.97 | 116 | 1.40 | 108 | 0.89 | 107 | 1.03 | 104 | 0.90 |
| $A_{1g}^1$ | 320 | 1.02 | 208 | 2.07 | 198 | 1.32 | 193 | 2.07 | 190 | 2.19 |
| $A_{1g}^2$ | 527 | 2.10 | 490 | 2.32 | 450 | 1.43[c], 2.00[d] | 427 | 1.77 | 408 | 1.47 |
| $E_g^2$ | 577 | 3.39 | 529 | 2.79 | 473 | 1.84[c], 3.40[d] | 437 | 3.33 | 418 | 3.23 |

[a] Estimated from Ref. 35 (22 GPa), [b] This work (11 GPa), [c] Ref. 28 (2 GPa), [d] Ref. 37 Hongo (3 GPa), [e] Ref. 30 (0 GPa), [f] Ref. 11 (0 GPa)

**Table S7**: Experimental wavenumbers at 0 GPa, $\omega_0$ (in cm$^{-1}$), for the Raman-active modes in several A-type $Ln_2O_3$ compounds. Our theoretical wavenumbers for Raman-active modes in A-type $Tb_2O_3$ at 0 GPa are noted in parenthesis.

| Sym. | $Y_2O_3$[a] $\omega_0$ | $Tb_2O_3$[b] $\omega_0$ | $\omega_0$ | $Sm_2O_3$[c] $\omega_0$ | $Nd_2O_3$[d] $\omega_0$ | $Pr_2O_3$[e] $\omega_0$ | $Ce_2O_3$[f] $\omega_0$ | $La_2O_3$[e] $\omega_0$ |
|---|---|---|---|---|---|---|---|---|
| $E_g^1$ | 153 | 99 | (102) | 106 | 107 | 104 | 103 | 104 |
| $A_{1g}^1$ | 298 | 183 | (185) | 195 | 193 | 187 | 189 | 190 |
| $A_{1g}^2$ | 481 | 465 | (455) | 447 | 427 | 406 |  | 400 |
| $E_g^2$ | 502 | 498 | (483) | 469 | 437 | 413 | 409 | 408 |
| $\Delta\omega$ | 21 | 33 |  | 22 | 10 | 7 |  | 8 |
| c/a | 1.617[g] | 1.577[g] |  | 1.572 | 1.567 | 1.558 | 1.557 | 1.555 |

[a] Estimated from Ref. 35, [b] Estimated from this work, [c] Estimated from Ref. 28, [d] Ref. 30, [e] Ref. 38, [f] Ref. 39, [g] Theoretical values at 0 GPa from Ref. 33.

**Table S8**: Theoretical frequencies $\omega_0$ (in cm$^{-1}$) at 0 GPa for the vibrational modes in A-type $Tb_2O_3$.

| Mode | $E_g^1$ | $A_{1g}^1$ | $E_u^1$ | $A_{2u}^1$ | $E_u^2$ | $A_{1g}^2$ | $A_{2u}^2$ | $E_g^2$ |
|---|---|---|---|---|---|---|---|---|
| $\omega_0$ | 101.7 | 184.8 | 193.8 | 221.6 | 437.1 | 455.0 | 467.1 | 483.0 |

# Calculation of the experimental and theoretical compressibility tensor of B-type $Tb_2O_3$ at different pressures

The isothermal compressibility tensor, $\beta_{ij}$, is a symmetric second rank tensor that relates the state of strain of a crystal to the change in pressure that induced the deformation **[41]**. The tensor coefficients for a monoclinic crystal with $b$ as the unique crystallographic axis are:

$$\beta_{ij} = \begin{pmatrix} \beta_{11} & 0 & \beta_{13} \\ 0 & \beta_{22} & 0 \\ \beta_{13} & 0 & \beta_{33} \end{pmatrix}$$

We have obtained the isothermal compressibility tensor coefficients for monoclinic B-type $Tb_2O_3$ at several pressures using the IRE (Institute of Radio Engineers) convention for the orthonormal basis for the tensor: $e_3 \| c$, $e_2 \| b^*$, $e_1 \| e_2 \times e_3$. The tensor has been obtained with the finite Eulerian approximation as implemented in the Win_Strain package **[42]**.

The change of the $\beta$ monoclinic angle (always perpendicular to the $b$ axis) with pressure implies that, in this monoclinic compound, the direction of the $a$ axis changes with pressure assuming both $b$ and $c$ axis constant. Furthermore, the departure of this monoclinic angle from 90° indicates that the direction of maximum compressibility is not exactly that of the $a$ axis. Therefore, in order to evaluate the direction of maximum compressibility as a function of pressure we have calculated and diagonalized the experimental and theoretical isothermal compressibility tensor, $\beta_{ij}$, at different pressures.

The experimental and theoretical elements of this tensor at different pressures are reported in **Tables S9** and **S10**, up to 11.0 GPa, where the directions of the maximum, intermediate and minimum compressibility and the values of the compressibility along those directions are given by the eigenvectors ($ev_i$, i=1-3) and eigenvalues ($\lambda_i$, i=1-3), respectively.

First of all, we have to note that there is a reasonable good agreement between the experimental and calculated axial compressibilities ($\beta_{ii}$ coefficients) at room pressure because $\beta_{11} > \beta_{33} > \beta_{22}$ in both cases. This result shows that the compressibility along the $a$-axis is greater than those to the $c$-axis and $b$-axis. A diagonalization of the $\beta_{ij}$ tensor at room pressure yields for our experiments the maximum, intermediate and minimum compressibilities $3.6(4) \cdot 10^{-3}$, $2.1(3) \cdot 10^{-3}$ and $1.9(6) \cdot 10^{-3}$ GPa$^{-1}$, respectively; whereas for the case of our calculations the obtained values for the compressibilities are $3.7(4) \cdot 10^{-3}$, $2.0(5) \cdot 10^{-3}$ and $1.8(3) \cdot 10^{-3}$ GPa$^{-1}$. These experimental (theoretical) results indicate that around 42% (43%) of the total compression at room pressure is being accommodated along the direction of maximum compressibility. Taking into account the eigenvector $ev_1$, the major compression direction at zero pressure occurs in the (010) plane at the given angle $\Psi$ (see **Tables S9 and S10**) relative to the $c$-axis (from $c$ to $a$) or equivalently at an angle $\theta$ relative to the *a*-axis (from *a* to *c*). In particular, the experimental major compression direction at room pressure is at $\theta=-19(5)°$ from the *a*-axis whereas for our calculations is at -17(3)° from the *a*-axis. The experimental direction of intermediate compressibility at room pressure, given by eigenvector $ev_2$, is in the (010) plane perpendicular to the direction of maximum compressibility, and the direction of minimum compressibility at room pressure, given by eigenvector $ev_3$, is along the *b* axis. On the other hand, in base of our *ab initio* calculations, the direction of intermediate compressibility at room pressure is along the *b*-axis, and the direction of minimum compressibility at room pressure is in the (0 1 0) plane perpendicular to the direction of maximum compressibility.

As regards the behavior of the experimental and theoretical compressibility tensor under pressure, it is found that $\beta_{11} > \beta_{33} > \beta_{22}$ is maintained as pressure increases. This result shows that a greater compressibility along the *a*-axis is found under pressure and that *b*-axis is the one that undergoes less compression. The experimental (theoretical) compressibility along the *a*-axis, $\beta_{11}$, increases slightly (does not change) under pressure. The compressibilities along *b*-axis and *c*-axis, $\beta_{22}$ and $\beta_{33}$, decrease as pressure increases.

The experimental (theoretical) maximum compressibility, $\lambda_1$, varies slightly (increases slightly) under compression. The intermediate and minimum compressibility, $\lambda_2$ and $\lambda_3$, decrease with pressure. Under compression, the experimental (theoretical) direction of maximum compressibility, $\theta$, approaches (moves away slightly) the *a*-axis. In both cases, the direction of maximum compressibility under pressure is always closer to the *a*-axis than to the *c*-axis. To conclude, the experimental and theoretical direction of intermediate compressibility under pressure is in the (010) plane perpendicular to the direction of maximum compressibility, and the direction of minimum compressibility is along the *b*-axis.

**Table S9.** Experimental isothermal compressibility tensor coefficients, $\beta_{ij}$, and their eigenvalues, $\lambda_i$, and eigenvectors, $ev_i$, for B-type $Tb_2O_3$ at several pressures. The results are given using the finite Eulerian method. The eigenvalues are given in decreasing value long a column.

| P(GPa) | 0.0 | 2.0 | 4.0 | 6.0 | 8.0 | 10.0 | 11.0 |
|---|---|---|---|---|---|---|---|
| $\beta_{11}$ ($10^{-3}$ $GPa^{-1}$) | 3.2(3) | 3.2(3) | 3.2(3) | 3.3(3) | 3.3(3) | 3.3(3) | 3.3(3) |
| $\beta_{22}$ ($10^{-3}$ $GPa^{-1}$) | 1.9(6) | 1.2(6) | 0.8(6) | 0.7(6) | 0.5(6) | 0.4(6) | 0.4(6) |
| $\beta_{33}$ ($10^{-3}$ $GPa^{-1}$) | 2.5(3) | 2.2(3) | 2.0(3) | 1.8(3) | 1.7(3) | 1.6(3) | 1.5(3) |
| $\beta_{13}$ ($10^{-3}$ $GPa^{-1}$) | -0.62(12) | -0.65(11) | -0.68(11) | -0.71(11) | -0.73(11) | -0.75(11) | -0.76(11) |
| $\lambda_1$ ($10^{-3}$ $GPa^{-1}$) | 3.6(4) | 3.5(3) | 3.5(3) | 3.6(3) | 3.6(3) | 3.6(3) | 3.6(3) |
| $ev_1$ ($\lambda_1$) | (0.87,0,-0.49) | (0.90,0,-0.44) | (0.91,0,-0.40) | (0.92,0,-0.38) | (0.93,0,-0.36) | (0.94,0,-0.35) | (0.94,0,-0.34) |
| $\lambda_2$ ($10^{-3}$ $GPa^{-1}$) | 2.1(3) | 1.9(3) | 1.7(3) | 1.5(3) | 1.4(3) | 1.3(3) | 1.2(3) |
| $ev_2$ ($\lambda_2$) | (0.49,0,0.87) | (0.44,0,0.90) | (0.40,0,0.91) | (0.38,0,0.92) | (0.36,0,0.93) | (0.35,0,0.94) | (0.34,0,0.94) |
| $\lambda_3$ ($10^{-3}$ $GPa^{-1}$) | 1.9(6) | 1.2(6) | 0.8(6) | 0.7(6) | 0.5(6) | 0.4(6) | 0.4(6) |
| $ev_3$ ($\lambda_3$) | (0,1,0) | (0,1,0) | (0,1,0) | (0,1,0) | (0,1,0) | (0,1,0) | (0,1,0) |
| $\Psi, \theta$ (°)[a] | 119(5), -19(5) | 116(5), -16(5) | 114(4), -14(4) | 112(4), -13(4) | 111(4), -12(4) | 110(3), -11(3) | 110(3), -11(3) |

[a] The major compression direction occurs in the (0 1 0) plane at the given angles $\Psi$ to the $c$-axis (from $c$ to $a$) and $\theta$ to the $a$-axis (from $a$ to $c$).

**Table S10.** Theoretical isothermal compressibility tensor coefficients, $\beta_{ij}$, and their eigenvalues, $\lambda_i$, and eigenvectors, $ev_i$, for B-type $Tb_2O_3$ at several pressures. The results are given using the finite Eulerian method. The eigenvalues are given in decreasing value along a column.

| P(GPa) | 0.0 | 2.0 | 4.0 | 6.0 | 8.0 | 10.0 | 11.0 |
|---|---|---|---|---|---|---|---|
| $\beta_{11}$ ($10^{-3}$ $GPa^{-1}$) | 3.3(3) | 3.3(3) | 3.3(3) | 3.3(3) | 3.3(3) | 3.3(3) | 3.3(3) |
| $\beta_{22}$ ($10^{-3}$ $GPa^{-1}$) | 2.0(4) | 1.6(4) | 1.2(4) | 1.0(4) | 0.8(4) | 0.7(4) | 0.7(4) |
| $\beta_{33}$ ($10^{-3}$ $GPa^{-1}$) | 2.2(3) | 2.08(22) | 2.02(22) | 1.96(22) | 1.90(22) | 1.85(21) | 1.82(18) |
| $\beta_{13}$ ($10^{-3}$ $GPa^{-1}$) | -0.77(9) | -0.81(9) | -0.85(9) | -0.91(9) | -0.97(9) | -1.05(10) | -1.09(8) |
| $\lambda_1$ ($10^{-3}$ $GPa^{-1}$) | 3.7(4) | 3.7(3) | 3.7(3) | 3.8(3) | 3.8(3) | 3.9(3) | 3.9(4) |
| $ev_1$ ($\lambda_1$) | (0.89,0,-0.46) | (0.89,0,-0.45) | (0.89,0,-0.45) | (0.89,0,-0.45) | (0.89,0,-0.45) | (0.89,0,-0.46) | (0.89,0,-0.47) |
| $\lambda_2$ ($10^{-3}$ $GPa^{-1}$) | 2.0(5) | 1.67(19) | 1.59(19) | 1.50(19) | 1.40(18) | 1.30(18) | 1.25(19) |
| $ev_2$ ($\lambda_2$) | (0,1,0) | (0.45,0,0.89) | (0.45,0,0.89) | (0.45,0,0.89) | (0.45,0,0.89) | (0.46,0,0.89) | (0.47,0,0.89) |
| $\lambda_3$ ($10^{-3}$ $GPa^{-1}$) | 1.8(3) | 1.6(4) | 1.2(4) | 1.0(4) | 0.8(4) | 0.7(4) | 0.7(4) |
| $ev_3$ ($\lambda_3$) | (0.46,0,0.89) | (0,1,0) | (0,1,0) | (0,1,0) | (0,1,0) | (0,1,0) | (0,1,0) |
| $\Psi, \theta$ (°)[a] | 117(3), -17(3) | 117(3), -17(3) | 117(3), -17(3) | 116.7(2.4), -16.9(2.4) | 117.0(2.2), -17.4(2.2) | 117.5(2.1), -18.0(2.1) | 117.7(1.8), -18.4(1.8) |

[a] The major compression direction occurs in the (0 1 0) plane at the given angles $\Psi$ to the $c$-axis (from $c$ to $a$) and $\theta$ to the $a$-axis (from $a$ to $c$).